\documentclass[%
preprint,
 amsmath,amssymb,
 aps,
 prx,
]{revtex4-2}
\usepackage{graphicx}%
\usepackage{multirow}%
\usepackage{amsmath,amssymb,amsfonts}%
\usepackage{amsthm}%
\usepackage{mathrsfs}%
\usepackage[title]{appendix}%
\usepackage{xcolor}%
\usepackage{textcomp}%
\usepackage{booktabs}%
\usepackage{algorithm}%
\usepackage{algorithmicx}%
\usepackage{algpseudocode}%
\usepackage{listings}%
\usepackage{psfrag}
\usepackage{dcolumn}
\usepackage{bm}
\usepackage{hyperref} 
\usepackage{xcolor}
\hypersetup{
    colorlinks,
    linkcolor={red!50!black},
    citecolor={blue!70!black},
    urlcolor={blue!90!black}
}
\usepackage{xr}
\begin{document}

\title[Article Title]{Three-dimensional Hard X-ray Ptychographic Reflectometry Imaging on Extended Mesoscopic Surface Structures}
\author{Peco Myint}
\thanks{These authors contributed equally to this work.}
\author{Ashish Tripathi}
\thanks{These authors contributed equally to this work.}
\author{Michael J. Wojcik}
\author{Junjing Deng}
\author{Mathew J. Cherukara}
\author{Nicholas Schwarz}
\author{Suresh Narayanan}
\author{Jin Wang}
\author{Miaoqi Chu}
\email{mqichu@anl.gov}
\thanks{Corresponding Author}
\author{Zhang Jiang}
\email{zjiang@anl.gov}
\thanks{Corresponding Author}\affiliation{X-ray Science Division, Advanced Photon Source, Argonne National Laboratory, 9700 S Cass Ave, Lemont, IL 60439, USA}

\begin{abstract}
{\noindent 
Many nano and quantum devices, with their sizes often spanning from millimeters down to sub-nanometer, have intricate low-dimensional, non-uniform, or hierarchical structures on surfaces and interfaces. Since their functionalities are dependent on these structures, high-resolution surface-sensitive characterization becomes imperative to gain a comprehensive understanding of the function-structure relationship. We thus developed hard X-ray ptychographic reflectometry imaging, a new technique that merges the high-resolution two-dimensional imaging capabilities of hard X-ray ptychography for extended objects, with the high-resolution depth profiling capabilities of X-ray reflectivity for layered structures. The synergy of these two methods fully leverages both amplitude and phase information from ptychography reconstruction to not only reveal surface topography and localized structures such as shapes and electron densities, but also yields statistical details such as interfacial roughness that is not readily accessible through coherent imaging solely. The hard X-ray ptychographic reflectometry imaging is well-suited for three-dimensional imaging of mesoscopic samples, particularly those comprising planar or layered nanostructures on opaque supports, and could also offer a high-resolution surface metrology and defect analysis on semiconductor devices such as integrated nanocircuits and lithographic photomasks for microchip fabrications.}
\end{abstract}

\maketitle
\section{Introduction}
Numerous functional devices consist of intricate low-dimensional, non-uniform, or hierarchical structures spanning from millimeters down to sub-nanometers on surfaces and interfaces. Examples of these mesoscale devices include layered or planar nanoelectronics~\cite{akinwande2014two,holler2017high,sangwan2020neuromorphic}, thin-film-based quantum dots~\cite{brown2014energy}, hetero-structured photovoltaic energy devices~\cite{nayak2019photovoltaic}, and hierarchically metamaterials~\cite{liu2010planar,xu2016planar}. Because their functions are closely tied to surface structures, it is essential to use high-resolution surface-sensitive characterization methods to understand the correlation between function and structure. Tools with nanometer-scaled spatial resolutions, such as scattering and imaging with electron or X-ray beam, are frequently employed. In comparison, X-rays excel at achieving greater penetration depth, non-destructive imaging, and offering a broader range of wavelengths compared to electrons. Notably, lensless coherent X-ray diffractive imaging (CXDI) technique~\cite{miao2015beyond} has seen significant advancements in recent decades, promising diffraction-limited resolutions of tens to a few nanometers. High-resolution reconstructions are achieved through iterative phase retrieval algorithms on over-sampled coherent images of dispersion or absorption contrast in a sample. However, CXDI requires the sample to be smaller than the incident probe and fully illuminated, limiting its applicability. In contrast, X-ray ptychography, an advancement of CXDI, offers the capability to image extended objects beyond the probe's field of view~\cite{rodenburg2007hard,pfeiffer2018x,guizar2021ptychography}. In ptychography, multiple CXDI images are captured through translational scans of the probe on the object while keeping sufficient overlap between adjacent scan positions. These images are reconstructed using ptychographic iterative engine (PIE) algorithms~\cite{rodenburg2019ptychography}, with updates made to both the probe and the object. This feature becomes particularly valuable as it enables the imaging of life-sized devices, even under in-operando conditions. The introduction of computed tomography further enhances these techniques, combining CXDI or ptychography from multiple projecting angles for a comprehensive 3D view of internal structures~\cite{dierolf2010ptychographic, holler2017high, holler2019three}. 

\begin{figure*}[t]%
\centering
\includegraphics[width=0.9\textwidth]{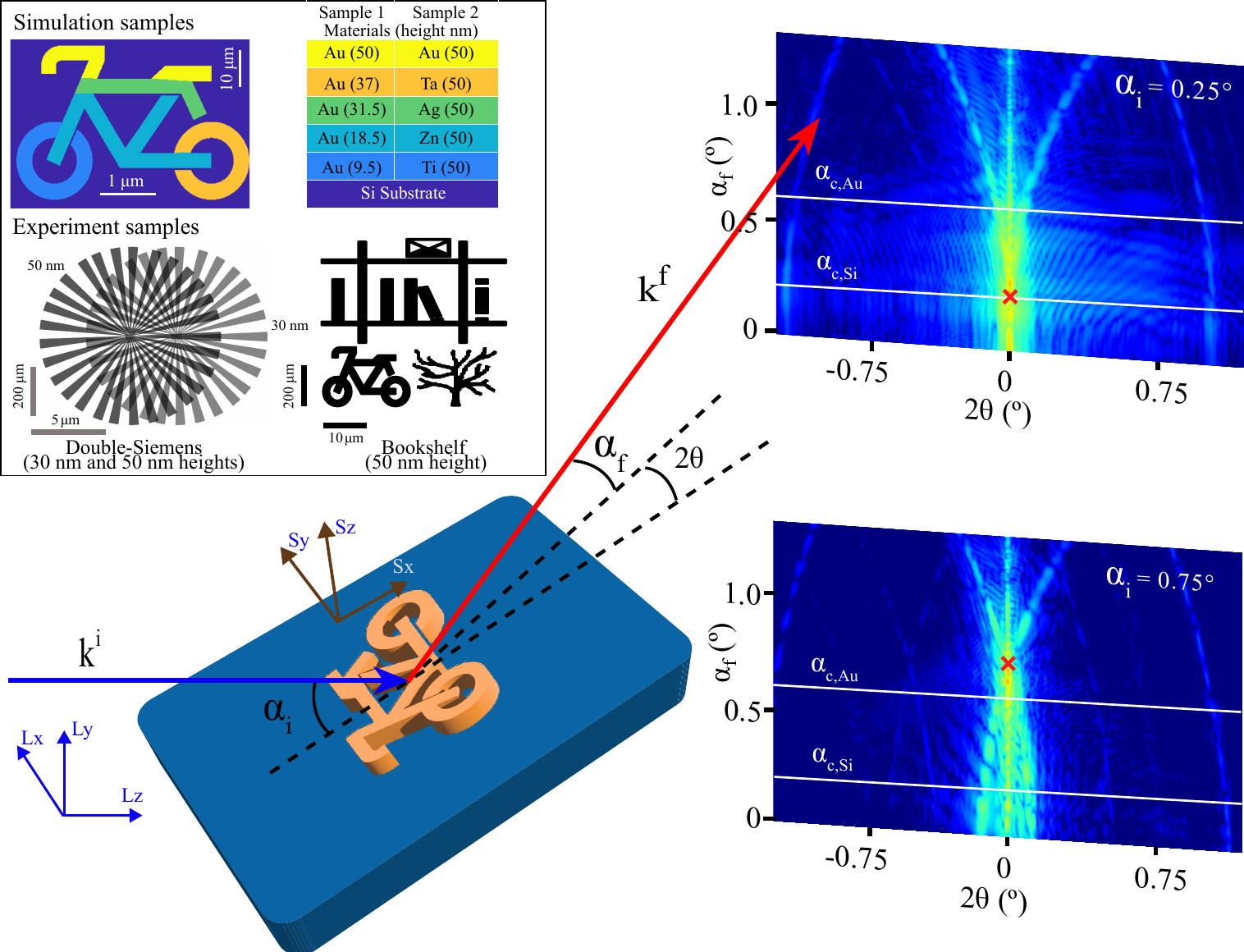}
\caption{Illustration hard X-ray ptychographic reflectometry imaging setup. A collimated or focused coherent beam impinges upon a pattern supported by a substrate. Incident angle $\alpha_i$ is labeled for the incident wave vector  $k^i$ along with exit angle $\alpha_f$ and in-plane azimuthal angle $2\theta$ for a scattered beam of wave vector $k^f$. The coordinate systems $(S_x,S_y,S_z)$ and $(L_x,L_y,L_z)$ correspond to the reference frames for the sample surface and the laboratory (i.e., incident beam), respectively. $S_xS_y$ is the plane where raster scanning is performed for ptychography. On the detector image, geometrical specular beam (GSB) (Supplementary \ref{supp:wave_vector_transfers}), where $\alpha_f=\alpha_i$ and $2\theta=0$, is indicated by a red cross. For 7.35~keV X-rays, the critical angles for total external reflection are $\alpha_{c,Si} = 0.245^{\circ}$ for silicon and $\alpha_{c,Au}=0.61^{\circ}$ for gold. These critical angles on the exit side are represented by horizontal white lines on the detector. Two illustrative coherent scattering images are provided from a section of the bicycle pattern for incident angles of $\alpha_i=0.25^{\circ}$ and $\alpha_i=0.75^{\circ}$. These images show the presence of multiple scatterings below the critical angles. The boxed inset displays the design parameters of the samples used in the simulation studies and the experiment.}\label{Fig:Imaging_geometry}
\end{figure*}

While these techniques are applicable to transmission~\cite{miao1999extending,thibault2008high} or Bragg diffraction~\cite{robinson2009coherent, clark2013ultrafast, ulvestad2017three,hruszkewycz2017high} geometries, there exists a demand for coherent imaging in reflection geometry to obtain high-resolution insights into non-crystalline mesoscopic structures on surfaces or in thin films. To address this need, we previously developed coherent surface scattering imaging (CSSI), where CXDI is performed on surface-supported isolated objects at grazing-incidence and a 2D in-plane view can be reconstructed single diffraction images. For a 3D view, the 3D reciprocal space is assembled from CSSI images taken at various incident angles and then reconstructed with 3D phase retrieval CXDI algorithms~\cite{sun2012three}. However, this method cannot be applied to extended structures buried in thin films or on impenetrable substrates. Here, we introduce three-dimensional hard X-ray ptychographic reflectometry imaging (Fig.~\ref{Fig:Imaging_geometry}) for extended surface objects, where in-plane views reconstructed from 2D ptychography reconstruction at multiple incident angles are combined for depth analysis using X-ray reflectivity in order for information normal to the surface. Conventional X-ray reflectivity analyzes reflection of X-rays off a material's surface. While it offers very high-resolution insights into layered structures and interface morphologies~\cite{tolan1999x,daillant2008x}, this information is averaged across the surface plane and does not provide any planar resolution. However, by harnessing high-resolution 2D imaging capabilities of ptychography and the precise 1D depth profiling offered in reflectometry, hard X-ray ptychographic reflectometry imaging can construct a complete 3D view of surface structures with nanometer resolutions over mesoscopic length scales. Therefore, it has the potential to apply to a variety of functional devices of planar or layered structures, including semiconducting devices such as integrated nanocircuits and reflective lithographical photomasks for microchip fabrications.

Recently, extreme ultraviolet (EUV) light (wavelength of about 10-100~nm) has been applied in reflective conditions to offer chemical sensitivity for surface structures~\cite{tanksalvala2021nondestructive}, albeit at angles far above grazing angles due to the EUV's relatively long wavelengths. Soft X-rays (wavelength of about 1-10~nm) were also demonstrated to image surface patterns in the holography mode~\cite{roy2011lensless}. Using coherent hard X-rays (energy $>6$~keV or wavelength $<0.2$~nm) for CXDI and ptychography at grazing-incident angles, on the other hand, not only enables enhanced spatial resolution due to much shorter wavelengths of hard X-rays and bright coherent flux at new-generation synchrotron facilities, but also facilitates deeper penetration beneath the surface to well beyond $1$~$\mu$m. Furthermore, it affords a dynamic range of $2-3$ orders of magnitude broader than EUV and soft X-rays for controlling penetration depth and surface sensitivity through adjustments in incident angles (Supplementary \ref{supp:penetration_depth}). In hard X-ray ptychographic reflectometry imaging, instead of solely measuring the intensity of the specular beam as in a conventional X-ray reflectivity, coherently scattered images are recorded at each incident angle (Fig.~\ref{Fig:Imaging_geometry}). In reflection geometry, amplitude contrast results from changes in reflectivity across different areas, whereas phase contrast primarily reveals height variations and provides a representation of the surface topography. Ptychography at each incident angle is reconstructed using phase-retrieval algorithms to produce a 2D projected view. These views at multiple incident angles are combined for a 3D view, akin to computerized tomography~\cite{withers2021x}. In this work, the layered structural information such as height, electron density, and roughness along the normal direction at various in-plane locations is obtained via 1D profile modeling. By uniting parameter-less ptychographic imaging with model-dependent reflectometry, this method becomes particularly effective for supported meso-scaled planar samples featuring highly asymmetric dimensions, where relevant structural sizes in the plane are often substantially larger than in the normal direction. As a variant of coherent diffractive imaging with the advantage of obviating imaging lenses, ptychographic reflectometry imaging can in principle achieve diffraction-limited transverse resolutions, akin to CXDI and ptychography in the transmission geometry. In this work, a transverse ($S_y$) resolution surpassing 34~nm and a normal ($S_z$) resolution of 2.4~nm were obtained. Because of the grazing-incidence geometry, the forward ($S_x$) resolution in the surface plane was approximately 2.4~$\mu$m, but it can be greatly improved through in-plane azimuthal rotations. 

Herein, we illustrate the principle of hard X-ray ptychographic reflectometry imaging and provide practical demonstrations using simulated and experimental examples. As an emerging non-destructive method for surface characterization, there exists substantial scope for further enhancement and innovation in its application, experimental implementations, and data analysis methodologies. For instance, dynamical scattering, often undesired in scattering and imaging experiments, possesses the potential to enable single-exposure 3D imaging~\cite{chu2023} if more advanced algorithms are developed, particularly when operating below critical angles for total external reflection. Furthermore, ptychographic reflectometry imaging can be combined with other techniques to broaden its capabilities. For instance, chemical sensitivity can be enhanced by using resonant energies. In addition, by substituting reflectometry with crystal truncation rod (CTR) scans~\cite{robinson1986crystal,kaganer2007crystal}, atomic structures at crystalline surfaces could be imaged with high in-plane resolutions.

\begin{figure*}[t]%
\centering
\includegraphics[width=0.9\textwidth]{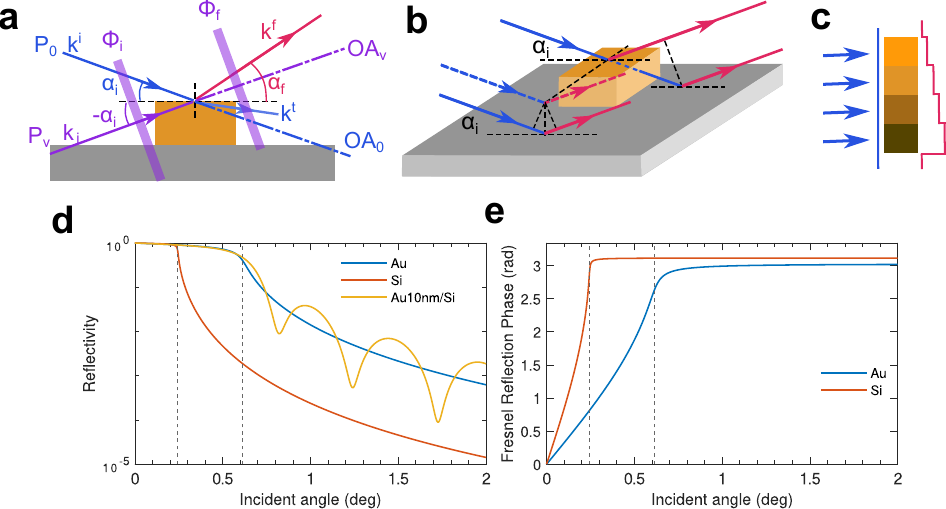}
\caption{Principle of hard X-ray ptychographic reflectometry imaging. (a) illustrates the relationship between the virtual probe $P_v$ and the actual incident probe $P_0$. Coherent diffractive imaging in reflection geometry is reconstructed with respect to the virtual optical axis $OA_v$ (purple solid line) rather than the real optical axis $OA_0$ (blue solid line). The wave-fronts of the virtual incident beam and the exit beam passing through the sample are represented by purple planes $\Phi_i$ and $\Phi_f$. (b) illustrates the origin of phase contrast, which is the path-length difference (PLD) from height variations in the plane. In contrast, (c) illustrates the mechanism of phase contrast in transmission geometry for hard X-ray imaging, where phase delays are induced by dispersion variations. For soft X-ray imaging, amplitude contrast is often employed due to large absorption variations. (d) displays calculated reflectivities for 7.35 keV X-rays from flat surfaces of bulk gold, bulk silicon, and a 10~nm gold layer on a silicon substrate, respectively, while (e) shows the phases of the Fresnel reflection coefficients (Eq.~\ref{Equ:Fresnel}) for gold and silicon. Vertical gray dotted lines indicate the critical angles for gold and silicon.}\label{Fig:Principle}
\end{figure*}

\section{Principles}
A unique characteristic of ptychography in the reflection geometry is the concept of ``virtual probe" $P_v$, which emerges from the reflection of the incident probe $P_0$ from an interface (Fig.~\ref{Fig:Principle}a). When an incident beam undergoes reflection, not only does its direction change, but its wave-front amplitude and phase are also altered according to the properties and structures of the underlying substances. When a probe is incident onto a flat surface of a bulk material at angle $\alpha_i$, the virtual probe can be envisioned as a beam impinging the surface at angle $-\alpha_i$. Its value is determined by the product of the real probe $P_0$ and the Fresnel reflection coefficient $r_F$, the latter given by 
\begin{equation}\label{Equ:Fresnel}
r_F(\alpha_i) = \frac{k^i_z - k^t_z}{k^i_z + k^t_z},
\end{equation}
where $k^i_z = k\sin\alpha_i$ and $k^t_z=k\sqrt{n^2-\cos^2{\alpha_i}}$ are, respectively, the normal components of the incident wave vector $k^i$ above and the transmitted wave vector $k^t$ below the interface. $n$ is the refractive index. Fig.~\ref{Fig:Principle}d and e show the calculated modulus-squared amplitude $R_F(\alpha_i)=|r_F(\alpha_i)|^2$ (often referred to as Fresnel reflectivity) and phase of Fresnel reflection coefficient for flat gold and silicon at the energy 7.35~keV. In case of layered planar films, the magnitude of the virtual probe is related to the specular reflectivity that can be calculated using Parratt's recursive method~\cite{parratt1954surface} as employed in conventional X-ray reflectivity analysis~\cite{tolan1999x,daillant2008x} and shown in Fig.~\ref{Fig:Principle}d for a 10~nm gold layer on silicon. If a surface with in-plane inhomogeneities (e.g. electron density variations) is scanned as in a ptychography, ptychographic reflectometry imaging  will unveil the depth dependence of the in-plane structures through profiling analysis on the in-plane dependent reflectivity
\begin{equation}\label{Equ:reflectivity}
    R(\alpha_i,S_x,S_y) = \left|\frac{P_v(\alpha_i)T(\alpha_i,S_x,S_y)}{P_0}\right|^2.
\end{equation}
Here the virtual probe $P_v(\alpha_i)$ remains fixed for all scanning positions during the ptychography reconstruction for a given incident angle, and $T(\alpha_i,S_x,S_y)$ is the reconstructed transfer function of the object projected along the virtual optical axis $OA_v$. At a fixed incident angle, $T(S_x,S_y)$ represents the ability of reflecting X-rays at in-plane location $(S_x,S_y)$. With X-ray reflectivity analysis on a desired in-plane region of interest (ROI), the amplitudes reconstructed from ptychography can give the structural information in the normal direction averaged in the ROI, implying a 3D imaging capability. In addition, due to the introduction of the virtual probe, the reconstructed view of the surface object resembles that of coherent diffractive imaging in the transmission geometry, especially in the small-angle scattering regime. Consequently, the remapping of reciprocal space is not required before the 2D ptychography reconstruction (Supplementary \ref{supp:wave_vector_transfers}). 

\begin{figure*}[t]%
\centering
\includegraphics[width=1\textwidth]{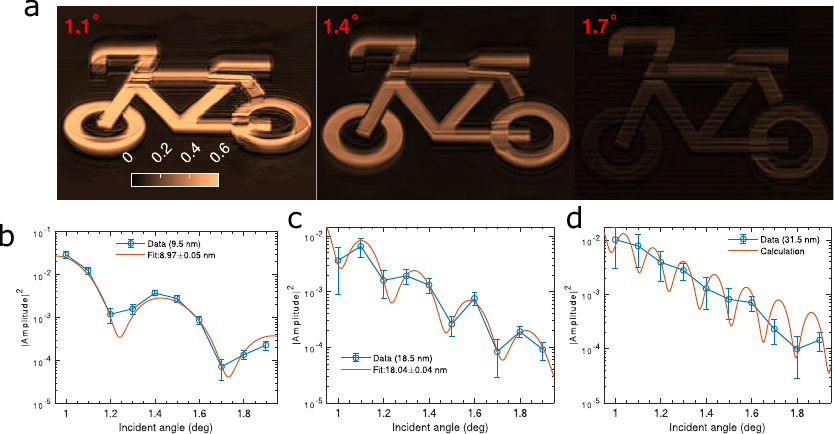}
\caption{Reconstructed amplitudes from sample \#1 of various heights. (a) shows the amplitudes of the reconstructed transfer function when viewed along the virtual optical axis, for three incident angles $1.1^\circ$,  $1.4^\circ$, and $1.7^\circ$. Note that, for convenience, all reconstructed objects (amplitude and phase) in this work are displayed in an inverted orientation (Supplementary \ref{supp:wave_vector_transfers}). (b-d) are averaged modulus squared amplitudes for three segments of the bicycle pattern: (b) front wheel (9.5 nm tall by design), (c) lower frame (18.5 nm), and (d) upper frame (31.5 nm). Red lines in (b,c) represent the best fits to the Parratt's reflectivity model (Appendix~\ref{app:reflectivity}), while the red line in (d) corresponds to the calculated reflectivity based on designed height of 31.5~nm for the upper frame.}\label{Fig:sample1_amplitude}
\end{figure*}

To illustrate the principle of ptychographic reflectometry imaging  for extended objects, we first conducted ptychography simulations on two distinct samples. These simulations were reconstructed using a modified ptychography algorithm that accommodates the representation of the virtual optical axis (Appendix~\ref{app:ptycho} and Supplementary \ref{supp:ptycho} and \ref{supp:wave_vector_transfers}). Since dynamical effects become significant when the incident angle falls below the critical angle~\cite{chu2023} (Supplementary \ref{supp:dynamical_scat}) and require further development on imaging reconstruction algorithms, we have chosen to steer away from low incident angles in this study in order to minimize the complexities associated with the dynamical effects. Sample \#1 consisted of gold segments with various heights, and sample \#2 featured segments composed of various materials (Au, Ta, Ag, Zn, Ti) but of a uniform height of 50~nm. Both samples consist of a bicycle pattern supported on a silicon substrate, and their design parameters are depicted in the inset of Fig.~\ref{Fig:Imaging_geometry}. Multi-slicing Fresnel propagation method~\cite{myint2023multislice, li2017multislice} was utilized as the forward method to compute the ptychography (Supplementary \ref{supp:multislicing}). Throughout the reconstruction, the flux of the virtual probe was set to remain constant, allowing variations of amplitude to serve as indicators of each segment's ability to reflect the incident beam, as demonstrated in the reconstructed amplitudes $|T(\alpha_i,x,y)|$ in Fig.~\ref{Fig:sample1_amplitude}a. Furthermore, as the incident angle increased, the reflectivity displayed oscillations that were dependent on segment height. Kiessig fringes, commonly observed in conventional reflectivity from thin films~\cite{tolan1999x,daillant2008x}, were evident on various segments. For instance, reflectivities (Eq.~\ref{Equ:reflectivity}) of the front wheel segment (9.5~nm tall by design and Fig.~\ref{Fig:sample1_amplitude}b) and the lower frame segment (18.5 nm, Fig.~\ref{Fig:sample1_amplitude}c) exhibited oscillating fringes. These fringes provided means to determine \textit{absolute} segment heights using Parratt's method for reflectivity analysis (Appendix~\ref{app:reflectivity}). It is noteworthy that substrate is not only a support material but also an integral part of the analysis. This is because the substrate acts as the reference and contributes to the coherent scattering by interfering with patterns on it. Hence, the amplitude reconstructed for the substrate (notice the discernible albeit weak substrate amplitude in Fig.~\ref{Fig:sample1_amplitude}) can be used to assess its localized information such as roughness and electron density, and its phase can serve as a reference to determine relative heights of other segments, as discussed below and in Supplementary \ref{supp:result_sam_2}.

A localized in-plane ROI needs to be defined before 1D profiling. The choice of its dimensions depends on the desired in-plane resolution. While the lower boundary of ROI is given by the best in-plane resolution from ptychography reconstruction (to be discussed later), its upper boundary should remain within the area where the structures in the normal direction exhibit homogeneity in the plane. In many cases, it can be determined by simultaneously examining the reconstructed amplitude and phase, and registering the reconstructed images obtained from various incident angles. Here as a straightforward demonstration, the entire area of each segment was used for its ROI averaging. In the 1D profiling, layer height and interface roughness were the only modeling parameters. Although a slight deviation was observed in the fitted height compared to the intended design value, it fell well within the resolution limits of the reflectivity analysis, which is determined by the maximum incident angle and given by $r_{Sz}=\lambda/[4\mathrm{max}(\sin\alpha_i)]=1.3$~nm for the simulation. In Parratt's method, interfacial roughness $\sigma$ manifests as an amplitude decay of the reflection and transmission coefficients of an interface. This is commonly described using the Nevot-Croce formula $\exp(-\frac{1}{2}k^2\sigma^2\sin^2\alpha_i)$~\cite{nevot1980caracterisation}, where $\sigma$ is root mean square (RMS) roughness. In our simulated study, we found a roughness of approximately 0.5~nm for each interface, for instance, $0.50\pm0.03$~nm and $0.55\pm0.07$~nm for the top and bottom surface of the front-wheel segment. Even though the simulation initially did not assume interfacial roughness, numerical errors during the voxel discretization when handling the sample's rotation effectively introduce rough interfaces that subsequently attenuate the amplitudes. The fitted values align with half of the voxel size of 1~nm used for modeling the cross-section of the sample (Supplementary \ref{supp:multislicing}). The hybrid technique of ptychography reflectometry provides a unique capability to extract localized information, in particular, the in-plane position-dependent local roughness. This capability cannot be achieved by either coherent imaging or traditional reflectivity methods alone. It is also noteworthy that successful application of the amplitude analysis requires adequate angular sampling of the fringes. As a counterexample, using a $0.1^\circ$ increment for sampling the incident angle is insufficient to extract the \textit{absolute} heights of the upper frame segment (31.5 nm by design) and other taller segments (Fig.~\ref{Fig:sample1_amplitude}d).

\begin{figure*}[t]%
\centering
\includegraphics[width=0.8\textwidth]{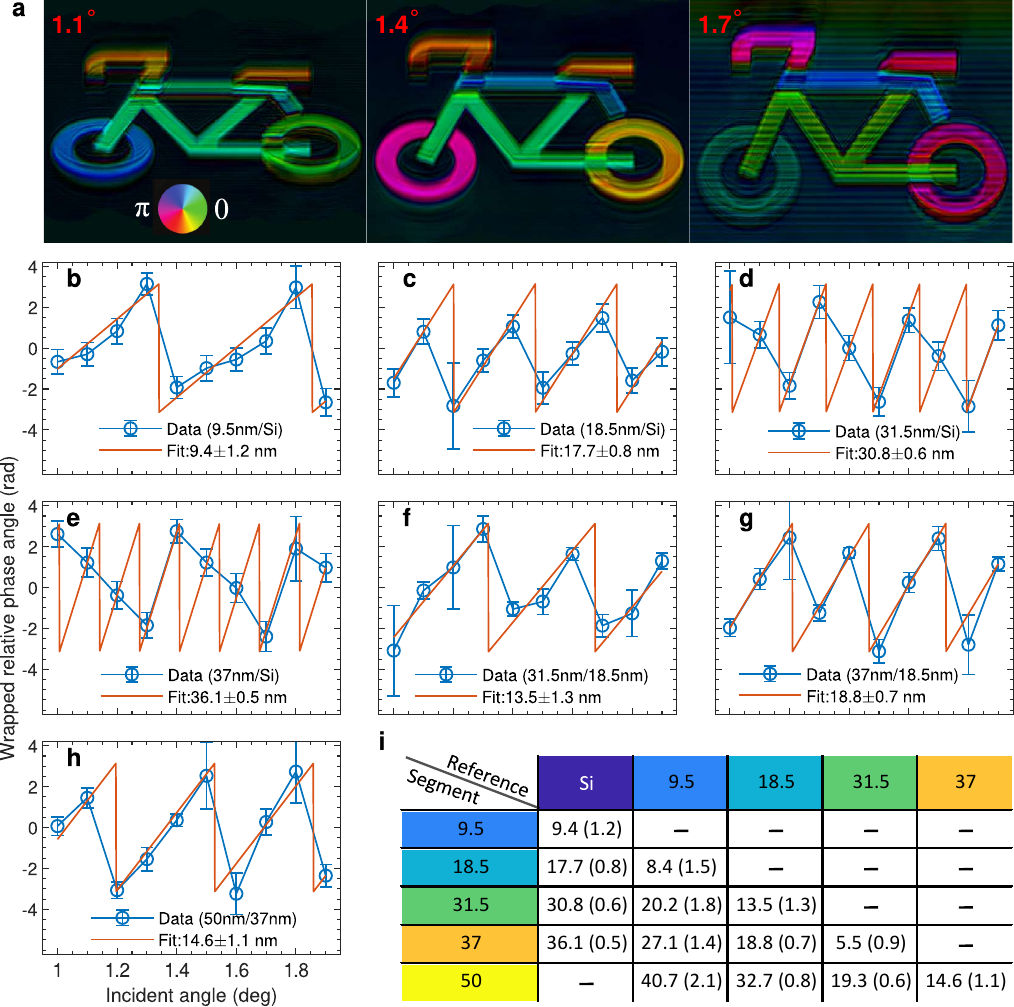}
\caption{Reconstructed phases from sample \#1 of various heights. (a) shows the HSV images of reconstructed transfer function when viewed along the virtual optical axis for three incident angles: $1.1^\circ$,  $1.4^\circ$, and $1.7^\circ$. HSV color represents the phase wrapped in the interval of $(-\pi,\pi]$, and brightness is the amplitude. (b-e) present wrapped relative phase delays averaged over each segment of the bicycle pattern using the substrate as the reference: (b) front wheel (9.5 nm tall by design), (c) lower frame (18.5 nm), (d) upper frame (31.5 nm), and (e) back wheel (37~nm). (f-h) are wrapped relative phase delays between segments as indicated in the legends. Red lines in (b-h) represent the best weighted robust (with the least-absolute-residual method) fits for the relative height using Eq.~\ref{Equ:phase}. (i) displays a table of relative heights (unit: nm) obtained from various pairs of segments and references (designated by color coding in the inset of Fig.~\ref{Fig:Imaging_geometry}). Values in parenthesis denote fitting uncertainties for a 95\% confidence interval. The unwrapped relative phases of all segment/reference pairs, along with the best fits, are displayed in Supplementary \ref{supp:unwrapped_relative_phase}.}\label{Fig:sample1_phase}
\end{figure*}

Fortunately, \textit{relative} heights of these under-sampled segments are encoded in the phases of the transfer function $T(\alpha_i,S_x,S_y)$, which is displayed in Fig.~\ref{Fig:sample1_phase}a as distinct phase contrasts from various segments on sample \#1. Phase contrast in the reflective geometry can be used as a valuable tool to map surface topography and it has two origins: phase shift upon Fresnel reflection (Fig.~\ref{Fig:Principle}e) and the path-length difference (PLD) from height variations in the plane (Fig.~\ref{Fig:Principle}b). When the incident angle is below the critical angle of total external reflection $\alpha_c$, X-rays do not penetrate through the material. Instead, they create evanescent waves near the surface~\cite{becker1983x}. This leads to a nearly 100\% reflection efficiency ($\alpha_i<\alpha_c$ in Fig.~\ref{Fig:Principle}d) as well as a gradually increased phase shift of the reflected beam from 0 to $\pi$ as the incident angle increases and approaches the critical angle (Fig.~\ref{Fig:Principle}e). This phase shift then remains nearly unchanged for angles above the critical angle. Therefore, for the angular range above the critical angle, PLD is the dominant factor determining the relative phase delays between different in-plane regions. Fig.~\ref{Fig:Principle}b illustrates that this phase delay can be expressed as
\begin{equation}\label{Equ:phase}
\Delta\varphi(\alpha_i) = 2\pi\left[\frac{2\Delta d}{\lambda}\sin\alpha_i + m\right],
\end{equation}
where $\Delta\varphi(\alpha_i)$ is the relative phase delay between segment and reference, $m$ is an integer, $\Delta d$ is the height difference, and $\lambda$ is the X-ray wavelength. Fig.~\ref{Fig:sample1_phase}b-e display the phases of the front wheel segment (9.5 nm), lower frame (18.5 nm), upper frame (31.5 nm), and back wheel (37 nm) with respect to that of the substrate. Red solid lines represent the best fits to the phase equation (Eq.~\ref{Equ:phase}), considering a phase wrapping to the interval $(-\pi,\pi]$. The fitted heights of the front wheel and the lower frame are consistent with both the designed values and their amplitude analysis. Interestingly, the relative phases of the upper frame (31.5 nm) and the back wheel (37 nm), whose absolute heights could not be determined via amplitude analysis, can reveal the relative heights against the substrate (Fig.~\ref{Fig:sample1_phase}d and e). However, it was unsuccessful to extract the height of the taller handlebar and seat segment (50 nm), if using the substrate as the phase reference. This is due to the uncertainty regarding the number of phase wrapping when applying Eq.~\ref{Equ:phase}. In other words, angular sampling rate is insufficient. For example, there are about 1.5 measured angles every $2\pi$ phase cycle for the back wheel segment (37 nm), barely feasible to extract the height (see unwrapped phase plots in Supplementary \ref{supp:unwrapped_relative_phase}). However, for the handlebar and seat segment (50 nm) with a sampling rate of $0.1^\circ$, there is only one measured angle every $2\pi$ phase cycle if using the substrate as the phase reference, making its height analysis unreliable and associated with very large uncertainties. One remedy is to increase the angular sampling rate to meet or exceed the required rate to resolve the height. Alternatively, segments of known heights can be used as the phase reference, as demonstrated in Fig.~\ref{Fig:sample1_phase}f-h. In this manner, heights of all segments were extracted, and cross-validated using various segment/reference pairs (Fig.~\ref{Fig:sample1_phase}i and Supplementary \ref{supp:unwrapped_relative_phase}). It is noteworthy that the relative phase analysis is insensitive to interfacial roughness. As a high-frequency statistical parameter, roughness does not significantly alter the overall PLD. Hence, it is more effectively resolved in the amplitude analysis as a contribution to the decay rate.

\begin{figure*}[t]%
\centering
\includegraphics[width=0.8\textwidth]{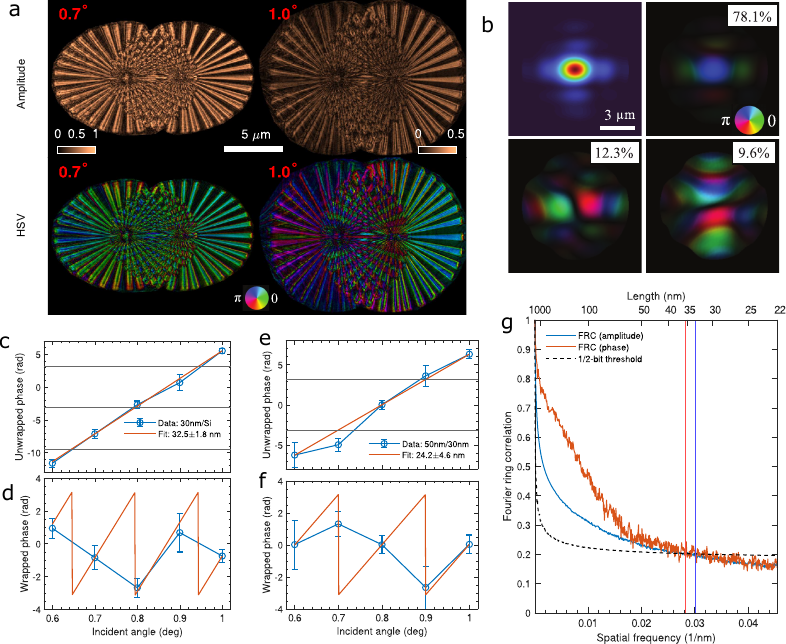}
\caption{Experimental ptychography reconstructions from a silicon supported double-Siemens pattern (inset in Fig.~\ref{Fig:Imaging_geometry} and Appendix~\ref{app:samples} for sample design). (a) shows the reconstructed amplitude (top row) and phase (HSV, bottom row) at $0.7^\circ$ and $1.0^\circ$. (b) shows the virtual probe: top-left panel is the total probe intensity $I_{\mathbf{v}} = \sum_{p=1}^3\vert \phi_p \vert^2$, and the other panels are HSV images of the dominant, secondary, and tertiary probe modes, each labeled with its respective mode occupancy.  (c) and (d) are the unwrapped and wrapped phases, respectively, using substrate as the reference for the height analysis of the 30~nm tall segment (right Siemens pattern). Similarly, (e) and (f) are the unwrapped and wrapped phases for the relative height analysis of the 50~nm tall segment (left Siemens pattern), using the 30~nm tall segment as the reference. Red lines in (c-f) are best robust fits to Eq.~\ref{Equ:phase}. Gray horizontal lines in (c) and (e) indicate points of phase wrapping for (d) and (f). (g) shows the Fourier ring correlation on the reconstructed amplitude (blue line) and phase (red line) at $0.8^\circ$. The black dotted line is the 1/2-bit threshold for the definition of resolution. The light blue and red vertical lines correspond to a transverse resolution of 33~nm and 35~nm from the FRC analysis on amplitude and phase, respectively. }\label{Fig:experiment}
\end{figure*}

\section{Experiment}
To experimentally validate the principle of the hard X-ray ptychographic reflectometry imaging from extended surface objects, we conducted experiments on a silicon supported gold sample featuring 30~nm and 50~nm tall Siemens patterns (inset of Fig.~\ref{Fig:Imaging_geometry} and Appendix~\ref{app:exp} and \ref{app:samples} for experiment conditions and sample details). Three spatial coherence probe modes (SCPM)~\cite{thibault2013reconstructing} were used during the reconstruction (Fig.~\ref{Fig:experiment}b). As the incident angle increases, the reconstructed amplitude reduces (top row in Fig.~\ref{Fig:experiment}a). As discussed previously, a sampling rate of 0.1$^\circ$ for the experiment (done with limited beam time access) is inadequate to extract the heights, and localized roughness as well, for this sample using the reflectivity method alone on the amplitude. However, we can utilize the relative phases. Bottom row in Fig.~\ref{Fig:experiment}a shows the reconstructed phases at two incident angles. The slight color variations observed over the large length scale along the vertical direction of the phase image (i.e. across the lateral dimension of 720~$\mu$m in the forward direction on the surface $S_x$) are believed to arise from the large-scale curvature or bending of the substrate, as this large-scale color variation is not obvious on other samples. For example, on the bookshelf sample (Appendix \ref{app:samples} and Supplementary \ref{supp:recon_bookshelf}) consisting of gold segments of identical height of 50~nm, reconstructed phase does not show pronounced variations across the entire sample. In the height analysis of the double-Siemens sample, the (vertical) central region was used (Fig.~\ref{Fig:experiment}c-f). Using the substrate nearby as the reference, $0.1^\circ$ angular sampling rate provides a good measurement of the height of the right-side Siemens pattern. The extracted height was $32.5\pm1.8$ nm. This is consistent with the design (30~nm thick gold and a 5~nm thick Ti adhesive layer). Similarly, when the right-side Siemens pattern was set as the reference, the left Siemens pattern was determined to 24.2~nm, still consistent with the designed values ($50-30=20$~nm), albeit with a large uncertainty. To further mitigate ambiguities and reduce the uncertainties, higher angular sampling rate and broader angular range as well as higher in-plane resolution (to be discussed below) would be advantageous. This can be achieved faster and efficiently with much brighter source at emerging diffracted-limited synchrotron facilities. Additionally, it has been noted that the edges of the reconstructed patterns display unevenness and fringes in both experimental and simulation samples. This occurrence is likely due to the shadow effect and multiple scatterings resulting from the relatively longer forward projection depth along the beam direction at grazing angles. Enhancements in this regard can be achieved through the adoption of advanced reconstruction algorithms currently undergoing further development, which entail multislicing wave propagation techniques.

Various well-established metrics are available for estimating spatial resolutions for the reconstruction, such as phase retrieval transfer function (PRTF)~\cite{steinbrener2010data} and Fourier ring correlation (FRC)~\cite{van2005fourier,vila2011characterization}. Fig.~\ref{Fig:experiment}g shows the Fourier ring correlation analysis on both amplitude and phase reconstructed for $0.8^\circ$. Using a signal-to-noise ratio (SNR) threshold of 1/2-bit, we determined transverse resolutions of 33~nm and 35~nm, corresponding to 1.5 and 1.6 pixels of the reconstructed object. We consider the averaged value, 34~nm, as the resolution $r_{Sy}$ along the $S_y$ direction. Due to the inclination angle $-\alpha_i$ that the virtual optical axis makes with the sample surface, the forward resolution in the surface plane (along $S_x$) becomes angle-dependent and can be calculated as $r_{Sx}=r_{Sy}/\sin\alpha_i$ (Supplementary \ref{supp:wave_vector_transfers}), which is 2.4~$\mu$m at $0.8^\circ$. It is worthwhile to note that increasing the incident angle improves the forward in-plane resolution; however, this improvement comes at a cost of reduced overall intensity and consequently a lower SNR as the incident angle increases. Lastly, for the resolution along the normal direction $S_z$, it was estimated using conventional definition in reflectivity analysis, which is 2.4~nm in this experiment. 

\section{Discussion}
Resolutions in hard X-ray ptychographic reflectometry imaging exhibit anisotropy due to the inherent properties of grazing-incidence geometry. The highest resolution is achieved along the sample's normal direction $S_z$, while the lowest resolution is along the in-plane forward direction $S_x$. To attain higher-resolution reconstructions in all directions, several enhancements can be implemented. These include expanding the detector's field of view and increasing incident angles. To achieve isotropic resolutions in the sample surface, one can conduct a tomographical scan involving azimuthal angles around the $S_z$ axis, akin to the approach of X-ray laminography~\cite{holler2019three}, albeit at grazing angles for ptychographic reflectometry imaging. Furthermore, the emerging diffraction-limited synchrotron facilities~\cite{eriksson2014diffraction}, such as the Advanced Photon Source Upgrade (APS-U), Extremely Brilliant Source of the European Synchrotron Radiation Facility (ESRF-EBS), Sirius, and MAX-IV, promise to deliver a significantly more intense beam with a coherent flux up to 100-1000 times greater than that of previous-generation synchrotrons. This advancement will not only enhance the resolution but also expedite experimental throughput, facilitating the examination of more realistic samples beyond those that have been demonstrated as a principle of concept in this work.

In this work, we adopted a high-angle configuration to minimize the impact of dynamical scattering in ptychographic reconstruction. It facilitates the application of conventional phase retrieval algorithms based on the kinematical approximation. Dynamical scattering phenomena are commonly observed in surface X-ray scattering experiments, particularly when dealing with grazing angles below the critical angle for external reflection. These phenomena arise from the perturbation of the substrate to the incident electric field. However, the kinematical approximation employed in most phase retrieval algorithms assumes an unchanged incident electric field and single scattering events. Consequently, it becomes inadequate in addressing dynamical scattering effects. Distorted wave Born approximation (DWBA) has shown its effectiveness in handling electric field perturbations in the surface normal direction~\cite{sinha1988x,jiang2011waveguide} and is widely utilized in grazing-incident small-angle X-ray scattering (GISAXS) analysis~\cite{renaud2009probing,hexemer2015advanced,li2016small} (Supplementary \ref{supp:dynamical_scat}). Yang and Sinha recently introduced a CXDI algorithm based on the DWBA framework assuming an in-plane homogeneous electric field and recovered the phase information from diffraction patterns generated by isolated surface objects in the grazing-incidence geometry~\cite{yang2023three}. When dealing with grazing-incidence coherent scattering imaging from surfaces featuring extended mesoscopic structures, perturbations to the in-plane electric field due to in-plane inhomogeneities often become significant. To address this, a finite-element DWBA (FE-DWBA) approach was recently developed~\cite{chu2023}. Nevertheless, the feasibility of using FE-DWBA framework in reconstructing ptychography remains to be explored. Additionally, DWBA assumes a plane-wave-like incident probe thus lacks the capability to account for other incident probes. To apply established ptychographic algorithms with only minimal adjustments for reflection geometry, we selected incident and exit angles in this study to exceed the critical angles of both the substrate and the materials constituting the sample. This approach effectively suppresses most dynamical scattering, and its mechanism can be analogously understood through classical illustrations in GISAXS analysis (Supplementary \ref{supp:dynamical_scat}). To be specific, a high incident angle ensures a clear separation of the reflective reciprocal space with respect to the virtual optical axis (two channels in Supplementary \ref{supp:Fig:DWBA}:c and d) from the transmitted reciprocal space with respect to the real optical axis (two channels in Supplementary \ref{supp:Fig:DWBA}:a and b). On the other hand, a high exit angle further minimizes the multiple beam events by reducing the contribution from the last channel in Supplementary \ref{supp:Fig:DWBA}:d. Consequently, the kinematical approximation becomes applicable. However, it is important to note that dynamical scattering as well as X-ray standing wave effects~\cite{pfeiffer2002two,zegenhagen2013x} at low angles carry significant information about the 3D structures and have the potential to facilitate high-resolution analysis, as noticed earlier~\cite{wang1992resonance,jiang2011waveguide,chu2023}. As an alternative to FE-DWBA for addressing dynamical scattering effects, multislicing Fresnel wave propagation has demonstrated accurate forward computation for coherent scattering from nanostructures on or beneath surfaces~\cite{myint2023multislice} (Supplementary \ref{supp:multislicing}). Phase retrieval using multislicing has been attempted in ptychography, albeit primarily in the transmission geometry and with a limited number of slices~\cite{maiden2012ptychographic,shimomura2015precession,tsai2016x,ozturk2018multi}. More advanced ptychographic reconstruction algorithms need to be developed to accommodate hundreds to thousands of slices required in multislicing when imaging surface samples or devices with mesoscopic structures. Strictly speaking, the use of high angles can mitigate the dynamical scattering but cannot entirely eradicate it, especially on objects of considerable longitudinal dimensions. Therefore, algorithms that can fully account for dynamical scattering not only improve the accuracy of ptychographic reconstruction but also harness the advantages of dynamical effects at grazing angles for high resolution and high surface sensitivity. 

In summary, we have introduced and demonstrated a hybrid technique for imaging surface structures using coherent hard X-rays on prototype samples. This approach combines the ptychography's 2D imaging capabilities of extended objects with the precise depth profiling capabilities of X-ray reflectivity. Consequently, it yields high-resolution 2D images of in-plane structures through parameter-less ptychographic reconstruction while extracting high-resolution information about structures in the normal direction through profile modeling with reflectivity analysis. The synergy of these two methods makes the most out of both amplitude and phase contrasts, thereby providing not only localized information regarding the features of surface structures such as height and shape, but also a statistical description, such as interface roughness, which can be otherwise challenging to obtain solely through ptychographic reconstruction. Moreover, the relative phase analysis proves effective in resolving thicker layers, even if the angular sampling rate might be insufficient for amplitude analysis. This capability has the potential to reduce data collection time, especially when topological information is the primary goal. Moreover, inherited from X-ray reflectivity, hard X-ray ptychographic reflectometry imaging has the sensitivity to electron density variations across various materials through profiling analysis that detects the shift of the critical angle and the decay envelope of the reflectivity. This capability was demonstrated with the simulated study on sample \#2, a bicycle pattern composed of various materials in different segments (Supplementary \ref{supp:result_sam_2}). In addition, the penetration depth of hard X-ray ptychographic reflectometry imaging can vary from a few nanometers to beyond tens of micrometers beneath the surface in the reflection geometry, eliminating the need for additional sample preparations, such as substrate thinning or sample milling, which are typically required for transmission-based coherent imaging methods. This aspect can be particularly advantageous when studying more realistic samples or devices with planar or layered nanostructures spanning mesoscopic length scales on the surface or in thin films supported on opaque substrates. Lastly, as a non-destructive technique, ptychographic reflectometry imaging may be integrated with other complementary methods. For example, one may carry out the experiment at a resonant energy of a specific element and use X-ray fluorescence to either improve elemental sensitivity or enhance structural resolution in thin films~\cite{sasaki1994fluorescent, jiang2020reconstruction}. If the incident energy is further scanned across the absorption edge of specific chemicals, additional contrasts can be induced to facilitate the mapping of these chemicals. Furthermore, ptychographic reflectometry imaging can be modified for crystal truncation rod scans, providing conventional CTR measurements~\cite{robinson1986crystal,kaganer2007crystal, zhu2015ptychographic} with in-plane resolution for visualizing atomic structures at surfaces. This can be particularly valuable for studying oxidation, epitaxial growth, and adsorption phenomena on laterally heterogeneous crystalline surfaces.

\section*{Acknowledgments}
This research used resources of the Advanced Photon Source and the Center for Nanoscale Materials, U.S. Department of Energy (DOE) Office of Science user facilities at Argonne National Laboratory and is based on research supported by the U.S. DOE Office of Science - Basic Energy Sciences, under Contract No. DE-AC02-06CH11357.  P.M., A.T., and Z.J. are also supported by the DOE Early Career Research Program. We thank Jongwoo Kim and Pice Chen for assistance on bookshelf sample fabrication, Daniel Ching and Viktor Nikitin for helpful discussions on data analysis, and Ray Ziegler for technical assistance during the experiments. 

\appendix
\section{Experiment Setup}\label{app:exp}

The experiments were performed at beamline Sector 8-ID-I of the Advanced Photon Source, Argonne National Laboratory, using X-rays of energy $E=7.35$~keV (wavelength $\lambda=1.687$~\AA) and monochromaticity $\Delta E/E =0.01\%$. A slit-collimated beam was focused with a set of compound refractive lenses (CRL) to a $\sim2\times2$~$\mu$m$^2$ (V$\times$H FWHM) spot size at the sample position. The total incoming photon flux amounted to about 6 $\times 10^{9}$~photons/s on the sample. The control and scanning of the sample's position and orientation was achieved using a stack of 11 stages consisting of Huber and SmarAct stages. The coherent images were captured by a Lambda (X-Spectrum GmbH) area detector of pixel size $55\times55$~$\mu$m$^2$ that was placed 3.82~m downstream from sample and with a vacuum flight path in between. The ptychography experiment was performed by scanning the sample surface through a two-dimensional point-to-point scan. Scan positions were first generated in the lab frame using Fermat spiral sequence~\cite{huang2014optimization} having an averaged overlapping ratio of about 90\%. The scan positions were transformed into the sample frame and the sample surface was scanned accordingly using SmarAct nano-precision linear stages. 

\section{Samples}\label{app:samples}
Two experimental samples (namely double-Siemens and bookshelf sample) were measured in this study. They were fabricated using the electron-beam lithography (EBL) and electron-beam evaporation facilities at the Center for Nanoscale Materials, Argonne National Laboratory. To make gold layer for the bookshelf and the first Siemens pattern, a poly(methyl methacrylate) (PMMA) resist layer was spin-coated onto a silicon wafer. After the pattern was created by a JEOL EBL system, the exposed regions were washed away. This was followed by the deposition of a 5-nm Ti adhesive layer and then the gold layer using an e-beam evaporator. This process was repeated for the second Siemens pattern. Both Siemens patterns have lateral dimensions of $12\times720$~$\mu$m$^2$, with a 3~$\mu$m offset along the shorter dimension. One pattern is 30~nm tall and the other is 50~nm. The dimensions of the bookshelf pattern are $40\times800$~$\mu$m$^2$ in the plane and 50~nm tall. 

For the simulation study, both samples have a bicycle pattern with identical lateral dimensions of $4\times 40$~$\mu$m$^2$ but of various heights and materials for different segments as displayed in the inset of Fig.~\ref{Fig:Imaging_geometry}. The bicycle pattern was directly placed on silicon without an adhesive layer. See Supplementary \ref{supp:multislicing} for simulation details. 

\section{Parratt's method for reflectivity}\label{app:reflectivity}
The principle of Parratt's recursive method is based on the interference of X-rays reflected from interfaces within a layered structure~\cite{tolan1999x,daillant2008x}. When X-rays impinge on an interface, a portion of the beam is reflected, while some is refracted and thus penetrates into the adjacent layer. The reflected and transmitted beams within each layer interfere with each other. These processes occur at each interface and within each layer except the topmost substance (e.g. vacuum or air) and the bottom-most substance (i.e. substrate). Starting from the bottom-most substance and moving up through all the layers to the topmost substance, Parratt's method recursively calculates the total reflected and transmitted waves within each layer using complex Fresnel reflection and transmission coefficients at each interface. The last iteration gives the total reflectivity, which is used for the height and roughness analysis on the reconstructed amplitudes from various segments in this study. 

\section{Ptychography reconstruction}\label{app:ptycho}
We modified the well-known rPIE (regularized ptychographic iterative engine)~\cite{maiden2017further} so that the sample, spatial coherence probe modes (SCPMs), and ptychographic scan positions \cite{Tripathi:14} are updated using ``stochastic minibatch gradient descent" method~\cite{minibatch_grad_intro} (Supplementary \ref{supp:ptycho}). First, ten trials of the ptychographic reconstruction were attempted using the same guessed initializations. The variance of the results in these ten trials arises from the stochastic minibatch selection (using a uniform random number generator for the selection of scan positions) of the measurements for each update of the 2000 epochs in total. The reconstructions with the lowest final error metric value were then used as initializations (with a small random perturbation to the sample) for more trials. The reconstructions shown here are from the results of the second set of ten trials with the lowest final error metric values. In the experimental ptychography reconstruction, we used a view of $768\times768$ pixels, symmetrically centered around the geometric specular beam on the detector image. In-house Matlab codes were used for reconstructions in this work, and similar algorithms have also been implemented in TIKE~\cite{Tike}, a ptychography toolbox developed at the Advanced Photon Source, Argonne National Laboratory. To prepare the Fourier ring correlation (FRC) analysis for the resolution estimation, we partitioned the complete ptychographic diffraction dataset into two subsets based on even and odd scan position indices, which were then independently reconstructed. The result with the lowest error metric for each subset was used to computer FRC. 

\bibliography{CSSIptycho}

\begin{thebibliography}{60}%
\makeatletter
\providecommand \@ifxundefined [1]{%
 \@ifx{#1\undefined}
}%
\providecommand \@ifnum [1]{%
 \ifnum #1\expandafter \@firstoftwo
 \else \expandafter \@secondoftwo
 \fi
}%
\providecommand \@ifx [1]{%
 \ifx #1\expandafter \@firstoftwo
 \else \expandafter \@secondoftwo
 \fi
}%
\providecommand \natexlab [1]{#1}%
\providecommand \enquote  [1]{``#1''}%
\providecommand \bibnamefont  [1]{#1}%
\providecommand \bibfnamefont [1]{#1}%
\providecommand \citenamefont [1]{#1}%
\providecommand \href@noop [0]{\@secondoftwo}%
\providecommand \href [0]{\begingroup \@sanitize@url \@href}%
\providecommand \@href[1]{\@@startlink{#1}\@@href}%
\providecommand \@@href[1]{\endgroup#1\@@endlink}%
\providecommand \@sanitize@url [0]{\catcode `\\12\catcode `\$12\catcode
  `\&12\catcode `\#12\catcode `\^12\catcode `\_12\catcode `\%12\relax}%
\providecommand \@@startlink[1]{}%
\providecommand \@@endlink[0]{}%
\providecommand \url  [0]{\begingroup\@sanitize@url \@url }%
\providecommand \@url [1]{\endgroup\@href {#1}{\urlprefix }}%
\providecommand \urlprefix  [0]{URL }%
\providecommand \Eprint [0]{\href }%
\providecommand \doibase [0]{https://doi.org/}%
\providecommand \selectlanguage [0]{\@gobble}%
\providecommand \bibinfo  [0]{\@secondoftwo}%
\providecommand \bibfield  [0]{\@secondoftwo}%
\providecommand \translation [1]{[#1]}%
\providecommand \BibitemOpen [0]{}%
\providecommand \bibitemStop [0]{}%
\providecommand \bibitemNoStop [0]{.\EOS\space}%
\providecommand \EOS [0]{\spacefactor3000\relax}%
\providecommand \BibitemShut  [1]{\csname bibitem#1\endcsname}%
\let\auto@bib@innerbib\@empty
\bibitem [{\citenamefont {Akinwande}\ \emph {et~al.}(2014)\citenamefont
  {Akinwande}, \citenamefont {Petrone},\ and\ \citenamefont
  {Hone}}]{akinwande2014two}%
  \BibitemOpen
  \bibfield  {author} {\bibinfo {author} {\bibfnamefont {D.}~\bibnamefont
  {Akinwande}}, \bibinfo {author} {\bibfnamefont {N.}~\bibnamefont {Petrone}},\
  and\ \bibinfo {author} {\bibfnamefont {J.}~\bibnamefont {Hone}},\ }\bibfield
  {title} {\bibinfo {title} {Two-dimensional flexible nanoelectronics},\
  }\href@noop {} {\bibfield  {journal} {\bibinfo  {journal} {Nature
  communications}\ }\textbf {\bibinfo {volume} {5}},\ \bibinfo {pages} {5678}
  (\bibinfo {year} {2014})}\BibitemShut {NoStop}%
\bibitem [{\citenamefont {Holler}\ \emph {et~al.}(2017)\citenamefont {Holler},
  \citenamefont {Guizar-Sicairos}, \citenamefont {Tsai}, \citenamefont
  {Dinapoli}, \citenamefont {M{\"u}ller}, \citenamefont {Bunk}, \citenamefont
  {Raabe},\ and\ \citenamefont {Aeppli}}]{holler2017high}%
  \BibitemOpen
  \bibfield  {author} {\bibinfo {author} {\bibfnamefont {M.}~\bibnamefont
  {Holler}}, \bibinfo {author} {\bibfnamefont {M.}~\bibnamefont
  {Guizar-Sicairos}}, \bibinfo {author} {\bibfnamefont {E.~H.}\ \bibnamefont
  {Tsai}}, \bibinfo {author} {\bibfnamefont {R.}~\bibnamefont {Dinapoli}},
  \bibinfo {author} {\bibfnamefont {E.}~\bibnamefont {M{\"u}ller}}, \bibinfo
  {author} {\bibfnamefont {O.}~\bibnamefont {Bunk}}, \bibinfo {author}
  {\bibfnamefont {J.}~\bibnamefont {Raabe}},\ and\ \bibinfo {author}
  {\bibfnamefont {G.}~\bibnamefont {Aeppli}},\ }\bibfield  {title} {\bibinfo
  {title} {High-resolution non-destructive three-dimensional imaging of
  integrated circuits},\ }\href@noop {} {\bibfield  {journal} {\bibinfo
  {journal} {Nature}\ }\textbf {\bibinfo {volume} {543}},\ \bibinfo {pages}
  {402} (\bibinfo {year} {2017})}\BibitemShut {NoStop}%
\bibitem [{\citenamefont {Sangwan}\ and\ \citenamefont
  {Hersam}(2020)}]{sangwan2020neuromorphic}%
  \BibitemOpen
  \bibfield  {author} {\bibinfo {author} {\bibfnamefont {V.~K.}\ \bibnamefont
  {Sangwan}}\ and\ \bibinfo {author} {\bibfnamefont {M.~C.}\ \bibnamefont
  {Hersam}},\ }\bibfield  {title} {\bibinfo {title} {Neuromorphic
  nanoelectronic materials},\ }\href@noop {} {\bibfield  {journal} {\bibinfo
  {journal} {Nature nanotechnology}\ }\textbf {\bibinfo {volume} {15}},\
  \bibinfo {pages} {517} (\bibinfo {year} {2020})}\BibitemShut {NoStop}%
\bibitem [{\citenamefont {Brown}\ \emph {et~al.}(2014)\citenamefont {Brown},
  \citenamefont {Kim}, \citenamefont {Lunt}, \citenamefont {Zhao},
  \citenamefont {Bawendi}, \citenamefont {Grossman},\ and\ \citenamefont
  {Bulovic}}]{brown2014energy}%
  \BibitemOpen
  \bibfield  {author} {\bibinfo {author} {\bibfnamefont {P.~R.}\ \bibnamefont
  {Brown}}, \bibinfo {author} {\bibfnamefont {D.}~\bibnamefont {Kim}}, \bibinfo
  {author} {\bibfnamefont {R.~R.}\ \bibnamefont {Lunt}}, \bibinfo {author}
  {\bibfnamefont {N.}~\bibnamefont {Zhao}}, \bibinfo {author} {\bibfnamefont
  {M.~G.}\ \bibnamefont {Bawendi}}, \bibinfo {author} {\bibfnamefont {J.~C.}\
  \bibnamefont {Grossman}},\ and\ \bibinfo {author} {\bibfnamefont
  {V.}~\bibnamefont {Bulovic}},\ }\bibfield  {title} {\bibinfo {title} {Energy
  level modification in lead sulfide quantum dot thin films through ligand
  exchange},\ }\href@noop {} {\bibfield  {journal} {\bibinfo  {journal} {ACS
  nano}\ }\textbf {\bibinfo {volume} {8}},\ \bibinfo {pages} {5863} (\bibinfo
  {year} {2014})}\BibitemShut {NoStop}%
\bibitem [{\citenamefont {Nayak}\ \emph {et~al.}(2019)\citenamefont {Nayak},
  \citenamefont {Mahesh}, \citenamefont {Snaith},\ and\ \citenamefont
  {Cahen}}]{nayak2019photovoltaic}%
  \BibitemOpen
  \bibfield  {author} {\bibinfo {author} {\bibfnamefont {P.~K.}\ \bibnamefont
  {Nayak}}, \bibinfo {author} {\bibfnamefont {S.}~\bibnamefont {Mahesh}},
  \bibinfo {author} {\bibfnamefont {H.~J.}\ \bibnamefont {Snaith}},\ and\
  \bibinfo {author} {\bibfnamefont {D.}~\bibnamefont {Cahen}},\ }\bibfield
  {title} {\bibinfo {title} {Photovoltaic solar cell technologies: analysing
  the state of the art},\ }\href@noop {} {\bibfield  {journal} {\bibinfo
  {journal} {Nature Reviews Materials}\ }\textbf {\bibinfo {volume} {4}},\
  \bibinfo {pages} {269} (\bibinfo {year} {2019})}\BibitemShut {NoStop}%
\bibitem [{\citenamefont {Liu}\ \emph {et~al.}(2010)\citenamefont {Liu},
  \citenamefont {Weiss}, \citenamefont {Mesch}, \citenamefont {Langguth},
  \citenamefont {Eigenthaler}, \citenamefont {Hirscher}, \citenamefont
  {Sonnichsen},\ and\ \citenamefont {Giessen}}]{liu2010planar}%
  \BibitemOpen
  \bibfield  {author} {\bibinfo {author} {\bibfnamefont {N.}~\bibnamefont
  {Liu}}, \bibinfo {author} {\bibfnamefont {T.}~\bibnamefont {Weiss}}, \bibinfo
  {author} {\bibfnamefont {M.}~\bibnamefont {Mesch}}, \bibinfo {author}
  {\bibfnamefont {L.}~\bibnamefont {Langguth}}, \bibinfo {author}
  {\bibfnamefont {U.}~\bibnamefont {Eigenthaler}}, \bibinfo {author}
  {\bibfnamefont {M.}~\bibnamefont {Hirscher}}, \bibinfo {author}
  {\bibfnamefont {C.}~\bibnamefont {Sonnichsen}},\ and\ \bibinfo {author}
  {\bibfnamefont {H.}~\bibnamefont {Giessen}},\ }\bibfield  {title} {\bibinfo
  {title} {Planar metamaterial analogue of electromagnetically induced
  transparency for plasmonic sensing},\ }\href@noop {} {\bibfield  {journal}
  {\bibinfo  {journal} {Nano letters}\ }\textbf {\bibinfo {volume} {10}},\
  \bibinfo {pages} {1103} (\bibinfo {year} {2010})}\BibitemShut {NoStop}%
\bibitem [{\citenamefont {Xu}\ \emph {et~al.}(2016)\citenamefont {Xu},
  \citenamefont {Fu},\ and\ \citenamefont {Chen}}]{xu2016planar}%
  \BibitemOpen
  \bibfield  {author} {\bibinfo {author} {\bibfnamefont {Y.}~\bibnamefont
  {Xu}}, \bibinfo {author} {\bibfnamefont {Y.}~\bibnamefont {Fu}},\ and\
  \bibinfo {author} {\bibfnamefont {H.}~\bibnamefont {Chen}},\ }\bibfield
  {title} {\bibinfo {title} {Planar gradient metamaterials},\ }\href@noop {}
  {\bibfield  {journal} {\bibinfo  {journal} {Nature Reviews Materials}\
  }\textbf {\bibinfo {volume} {1}},\ \bibinfo {pages} {1} (\bibinfo {year}
  {2016})}\BibitemShut {NoStop}%
\bibitem [{\citenamefont {Miao}\ \emph {et~al.}(2015)\citenamefont {Miao},
  \citenamefont {Ishikawa}, \citenamefont {Robinson},\ and\ \citenamefont
  {Murnane}}]{miao2015beyond}%
  \BibitemOpen
  \bibfield  {author} {\bibinfo {author} {\bibfnamefont {J.}~\bibnamefont
  {Miao}}, \bibinfo {author} {\bibfnamefont {T.}~\bibnamefont {Ishikawa}},
  \bibinfo {author} {\bibfnamefont {I.~K.}\ \bibnamefont {Robinson}},\ and\
  \bibinfo {author} {\bibfnamefont {M.~M.}\ \bibnamefont {Murnane}},\
  }\bibfield  {title} {\bibinfo {title} {Beyond crystallography: Diffractive
  imaging using coherent x-ray light sources},\ }\href@noop {} {\bibfield
  {journal} {\bibinfo  {journal} {Science}\ }\textbf {\bibinfo {volume}
  {348}},\ \bibinfo {pages} {530} (\bibinfo {year} {2015})}\BibitemShut
  {NoStop}%
\bibitem [{\citenamefont {Rodenburg}\ \emph {et~al.}(2007)\citenamefont
  {Rodenburg}, \citenamefont {Hurst}, \citenamefont {Cullis}, \citenamefont
  {Dobson}, \citenamefont {Pfeiffer}, \citenamefont {Bunk}, \citenamefont
  {David}, \citenamefont {Jefimovs},\ and\ \citenamefont
  {Johnson}}]{rodenburg2007hard}%
  \BibitemOpen
  \bibfield  {author} {\bibinfo {author} {\bibfnamefont {J.~M.}\ \bibnamefont
  {Rodenburg}}, \bibinfo {author} {\bibfnamefont {A.}~\bibnamefont {Hurst}},
  \bibinfo {author} {\bibfnamefont {A.~G.}\ \bibnamefont {Cullis}}, \bibinfo
  {author} {\bibfnamefont {B.~R.}\ \bibnamefont {Dobson}}, \bibinfo {author}
  {\bibfnamefont {F.}~\bibnamefont {Pfeiffer}}, \bibinfo {author}
  {\bibfnamefont {O.}~\bibnamefont {Bunk}}, \bibinfo {author} {\bibfnamefont
  {C.}~\bibnamefont {David}}, \bibinfo {author} {\bibfnamefont
  {K.}~\bibnamefont {Jefimovs}},\ and\ \bibinfo {author} {\bibfnamefont
  {I.}~\bibnamefont {Johnson}},\ }\bibfield  {title} {\bibinfo {title}
  {Hard-x-ray lensless imaging of extended objects},\ }\href@noop {} {\bibfield
   {journal} {\bibinfo  {journal} {Physical review letters}\ }\textbf {\bibinfo
  {volume} {98}},\ \bibinfo {pages} {034801} (\bibinfo {year}
  {2007})}\BibitemShut {NoStop}%
\bibitem [{\citenamefont {Pfeiffer}(2018)}]{pfeiffer2018x}%
  \BibitemOpen
  \bibfield  {author} {\bibinfo {author} {\bibfnamefont {F.}~\bibnamefont
  {Pfeiffer}},\ }\bibfield  {title} {\bibinfo {title} {X-ray ptychography},\
  }\href@noop {} {\bibfield  {journal} {\bibinfo  {journal} {Nature Photonics}\
  }\textbf {\bibinfo {volume} {12}},\ \bibinfo {pages} {9} (\bibinfo {year}
  {2018})}\BibitemShut {NoStop}%
\bibitem [{\citenamefont {Guizar-Sicairos}\ and\ \citenamefont
  {Thibault}(2021)}]{guizar2021ptychography}%
  \BibitemOpen
  \bibfield  {author} {\bibinfo {author} {\bibfnamefont {M.}~\bibnamefont
  {Guizar-Sicairos}}\ and\ \bibinfo {author} {\bibfnamefont {P.}~\bibnamefont
  {Thibault}},\ }\bibfield  {title} {\bibinfo {title} {Ptychography: A solution
  to the phase problem},\ }\href@noop {} {\bibfield  {journal} {\bibinfo
  {journal} {Physics Today}\ }\textbf {\bibinfo {volume} {74}},\ \bibinfo
  {pages} {42} (\bibinfo {year} {2021})}\BibitemShut {NoStop}%
\bibitem [{\citenamefont {Rodenburg}\ and\ \citenamefont
  {Maiden}(2019)}]{rodenburg2019ptychography}%
  \BibitemOpen
  \bibfield  {author} {\bibinfo {author} {\bibfnamefont {J.}~\bibnamefont
  {Rodenburg}}\ and\ \bibinfo {author} {\bibfnamefont {A.}~\bibnamefont
  {Maiden}},\ }\bibfield  {title} {\bibinfo {title} {Ptychography},\
  }\href@noop {} {\bibfield  {journal} {\bibinfo  {journal} {Springer Handbook
  of Microscopy}\ ,\ \bibinfo {pages} {819}} (\bibinfo {year}
  {2019})}\BibitemShut {NoStop}%
\bibitem [{\citenamefont {Dierolf}\ \emph {et~al.}(2010)\citenamefont
  {Dierolf}, \citenamefont {Menzel}, \citenamefont {Thibault}, \citenamefont
  {Schneider}, \citenamefont {Kewish}, \citenamefont {Wepf}, \citenamefont
  {Bunk},\ and\ \citenamefont {Pfeiffer}}]{dierolf2010ptychographic}%
  \BibitemOpen
  \bibfield  {author} {\bibinfo {author} {\bibfnamefont {M.}~\bibnamefont
  {Dierolf}}, \bibinfo {author} {\bibfnamefont {A.}~\bibnamefont {Menzel}},
  \bibinfo {author} {\bibfnamefont {P.}~\bibnamefont {Thibault}}, \bibinfo
  {author} {\bibfnamefont {P.}~\bibnamefont {Schneider}}, \bibinfo {author}
  {\bibfnamefont {C.~M.}\ \bibnamefont {Kewish}}, \bibinfo {author}
  {\bibfnamefont {R.}~\bibnamefont {Wepf}}, \bibinfo {author} {\bibfnamefont
  {O.}~\bibnamefont {Bunk}},\ and\ \bibinfo {author} {\bibfnamefont
  {F.}~\bibnamefont {Pfeiffer}},\ }\bibfield  {title} {\bibinfo {title}
  {Ptychographic x-ray computed tomography at the nanoscale},\ }\href@noop {}
  {\bibfield  {journal} {\bibinfo  {journal} {Nature}\ }\textbf {\bibinfo
  {volume} {467}},\ \bibinfo {pages} {436} (\bibinfo {year}
  {2010})}\BibitemShut {NoStop}%
\bibitem [{\citenamefont {Holler}\ \emph {et~al.}(2019)\citenamefont {Holler},
  \citenamefont {Odstrcil}, \citenamefont {Guizar-Sicairos}, \citenamefont
  {Lebugle}, \citenamefont {M{\"u}ller}, \citenamefont {Finizio}, \citenamefont
  {Tinti}, \citenamefont {David}, \citenamefont {Zusman}, \citenamefont
  {Unglaub} \emph {et~al.}}]{holler2019three}%
  \BibitemOpen
  \bibfield  {author} {\bibinfo {author} {\bibfnamefont {M.}~\bibnamefont
  {Holler}}, \bibinfo {author} {\bibfnamefont {M.}~\bibnamefont {Odstrcil}},
  \bibinfo {author} {\bibfnamefont {M.}~\bibnamefont {Guizar-Sicairos}},
  \bibinfo {author} {\bibfnamefont {M.}~\bibnamefont {Lebugle}}, \bibinfo
  {author} {\bibfnamefont {E.}~\bibnamefont {M{\"u}ller}}, \bibinfo {author}
  {\bibfnamefont {S.}~\bibnamefont {Finizio}}, \bibinfo {author} {\bibfnamefont
  {G.}~\bibnamefont {Tinti}}, \bibinfo {author} {\bibfnamefont
  {C.}~\bibnamefont {David}}, \bibinfo {author} {\bibfnamefont
  {J.}~\bibnamefont {Zusman}}, \bibinfo {author} {\bibfnamefont
  {W.}~\bibnamefont {Unglaub}}, \emph {et~al.},\ }\bibfield  {title} {\bibinfo
  {title} {Three-dimensional imaging of integrated circuits with macro-to
  nanoscale zoom},\ }\href@noop {} {\bibfield  {journal} {\bibinfo  {journal}
  {Nature Electronics}\ }\textbf {\bibinfo {volume} {2}},\ \bibinfo {pages}
  {464} (\bibinfo {year} {2019})}\BibitemShut {NoStop}%
\bibitem [{\citenamefont {Miao}\ \emph {et~al.}(1999)\citenamefont {Miao},
  \citenamefont {Charalambous}, \citenamefont {Kirz},\ and\ \citenamefont
  {Sayre}}]{miao1999extending}%
  \BibitemOpen
  \bibfield  {author} {\bibinfo {author} {\bibfnamefont {J.}~\bibnamefont
  {Miao}}, \bibinfo {author} {\bibfnamefont {P.}~\bibnamefont {Charalambous}},
  \bibinfo {author} {\bibfnamefont {J.}~\bibnamefont {Kirz}},\ and\ \bibinfo
  {author} {\bibfnamefont {D.}~\bibnamefont {Sayre}},\ }\bibfield  {title}
  {\bibinfo {title} {Extending the methodology of x-ray crystallography to
  allow imaging of micrometre-sized non-crystalline specimens},\ }\href@noop {}
  {\bibfield  {journal} {\bibinfo  {journal} {Nature}\ }\textbf {\bibinfo
  {volume} {400}},\ \bibinfo {pages} {342} (\bibinfo {year}
  {1999})}\BibitemShut {NoStop}%
\bibitem [{\citenamefont {Thibault}\ \emph {et~al.}(2008)\citenamefont
  {Thibault}, \citenamefont {Dierolf}, \citenamefont {Menzel}, \citenamefont
  {Bunk}, \citenamefont {David},\ and\ \citenamefont
  {Pfeiffer}}]{thibault2008high}%
  \BibitemOpen
  \bibfield  {author} {\bibinfo {author} {\bibfnamefont {P.}~\bibnamefont
  {Thibault}}, \bibinfo {author} {\bibfnamefont {M.}~\bibnamefont {Dierolf}},
  \bibinfo {author} {\bibfnamefont {A.}~\bibnamefont {Menzel}}, \bibinfo
  {author} {\bibfnamefont {O.}~\bibnamefont {Bunk}}, \bibinfo {author}
  {\bibfnamefont {C.}~\bibnamefont {David}},\ and\ \bibinfo {author}
  {\bibfnamefont {F.}~\bibnamefont {Pfeiffer}},\ }\bibfield  {title} {\bibinfo
  {title} {High-resolution scanning x-ray diffraction microscopy},\ }\href@noop
  {} {\bibfield  {journal} {\bibinfo  {journal} {Science}\ }\textbf {\bibinfo
  {volume} {321}},\ \bibinfo {pages} {379} (\bibinfo {year}
  {2008})}\BibitemShut {NoStop}%
\bibitem [{\citenamefont {Robinson}\ and\ \citenamefont
  {Harder}(2009)}]{robinson2009coherent}%
  \BibitemOpen
  \bibfield  {author} {\bibinfo {author} {\bibfnamefont {I.}~\bibnamefont
  {Robinson}}\ and\ \bibinfo {author} {\bibfnamefont {R.}~\bibnamefont
  {Harder}},\ }\bibfield  {title} {\bibinfo {title} {Coherent x-ray diffraction
  imaging of strain at the nanoscale},\ }\href@noop {} {\bibfield  {journal}
  {\bibinfo  {journal} {Nature materials}\ }\textbf {\bibinfo {volume} {8}},\
  \bibinfo {pages} {291} (\bibinfo {year} {2009})}\BibitemShut {NoStop}%
\bibitem [{\citenamefont {Clark}\ \emph {et~al.}(2013)\citenamefont {Clark},
  \citenamefont {Beitra}, \citenamefont {Xiong}, \citenamefont {Higginbotham},
  \citenamefont {Fritz}, \citenamefont {Lemke}, \citenamefont {Zhu},
  \citenamefont {Chollet}, \citenamefont {Williams}, \citenamefont
  {Messerschmidt} \emph {et~al.}}]{clark2013ultrafast}%
  \BibitemOpen
  \bibfield  {author} {\bibinfo {author} {\bibfnamefont {J.}~\bibnamefont
  {Clark}}, \bibinfo {author} {\bibfnamefont {L.}~\bibnamefont {Beitra}},
  \bibinfo {author} {\bibfnamefont {G.}~\bibnamefont {Xiong}}, \bibinfo
  {author} {\bibfnamefont {A.}~\bibnamefont {Higginbotham}}, \bibinfo {author}
  {\bibfnamefont {D.}~\bibnamefont {Fritz}}, \bibinfo {author} {\bibfnamefont
  {H.}~\bibnamefont {Lemke}}, \bibinfo {author} {\bibfnamefont
  {D.}~\bibnamefont {Zhu}}, \bibinfo {author} {\bibfnamefont {M.}~\bibnamefont
  {Chollet}}, \bibinfo {author} {\bibfnamefont {G.}~\bibnamefont {Williams}},
  \bibinfo {author} {\bibfnamefont {M.}~\bibnamefont {Messerschmidt}}, \emph
  {et~al.},\ }\bibfield  {title} {\bibinfo {title} {Ultrafast three-dimensional
  imaging of lattice dynamics in individual gold nanocrystals},\ }\href@noop {}
  {\bibfield  {journal} {\bibinfo  {journal} {Science}\ }\textbf {\bibinfo
  {volume} {341}},\ \bibinfo {pages} {56} (\bibinfo {year} {2013})}\BibitemShut
  {NoStop}%
\bibitem [{\citenamefont {Ulvestad}\ \emph {et~al.}(2017)\citenamefont
  {Ulvestad}, \citenamefont {Welland}, \citenamefont {Cha}, \citenamefont
  {Liu}, \citenamefont {Kim}, \citenamefont {Harder}, \citenamefont {Maxey},
  \citenamefont {Clark}, \citenamefont {Highland}, \citenamefont {You} \emph
  {et~al.}}]{ulvestad2017three}%
  \BibitemOpen
  \bibfield  {author} {\bibinfo {author} {\bibfnamefont {A.}~\bibnamefont
  {Ulvestad}}, \bibinfo {author} {\bibfnamefont {M.}~\bibnamefont {Welland}},
  \bibinfo {author} {\bibfnamefont {W.}~\bibnamefont {Cha}}, \bibinfo {author}
  {\bibfnamefont {Y.}~\bibnamefont {Liu}}, \bibinfo {author} {\bibfnamefont
  {J.}~\bibnamefont {Kim}}, \bibinfo {author} {\bibfnamefont {R.}~\bibnamefont
  {Harder}}, \bibinfo {author} {\bibfnamefont {E.}~\bibnamefont {Maxey}},
  \bibinfo {author} {\bibfnamefont {J.}~\bibnamefont {Clark}}, \bibinfo
  {author} {\bibfnamefont {M.}~\bibnamefont {Highland}}, \bibinfo {author}
  {\bibfnamefont {H.}~\bibnamefont {You}}, \emph {et~al.},\ }\bibfield  {title}
  {\bibinfo {title} {Three-dimensional imaging of dislocation dynamics during
  the hydriding phase transformation},\ }\href@noop {} {\bibfield  {journal}
  {\bibinfo  {journal} {Nature materials}\ }\textbf {\bibinfo {volume} {16}},\
  \bibinfo {pages} {565} (\bibinfo {year} {2017})}\BibitemShut {NoStop}%
\bibitem [{\citenamefont {Hruszkewycz}\ \emph {et~al.}(2017)\citenamefont
  {Hruszkewycz}, \citenamefont {Allain}, \citenamefont {Holt}, \citenamefont
  {Murray}, \citenamefont {Holt}, \citenamefont {Fuoss},\ and\ \citenamefont
  {Chamard}}]{hruszkewycz2017high}%
  \BibitemOpen
  \bibfield  {author} {\bibinfo {author} {\bibfnamefont {S.~O.}\ \bibnamefont
  {Hruszkewycz}}, \bibinfo {author} {\bibfnamefont {M.}~\bibnamefont {Allain}},
  \bibinfo {author} {\bibfnamefont {M.~V.}\ \bibnamefont {Holt}}, \bibinfo
  {author} {\bibfnamefont {C.~E.}\ \bibnamefont {Murray}}, \bibinfo {author}
  {\bibfnamefont {J.}~\bibnamefont {Holt}}, \bibinfo {author} {\bibfnamefont
  {P.}~\bibnamefont {Fuoss}},\ and\ \bibinfo {author} {\bibfnamefont
  {V.}~\bibnamefont {Chamard}},\ }\bibfield  {title} {\bibinfo {title}
  {High-resolution three-dimensional structural microscopy by single-angle
  bragg ptychography},\ }\href@noop {} {\bibfield  {journal} {\bibinfo
  {journal} {Nature materials}\ }\textbf {\bibinfo {volume} {16}},\ \bibinfo
  {pages} {244} (\bibinfo {year} {2017})}\BibitemShut {NoStop}%
\bibitem [{\citenamefont {Sun}\ \emph {et~al.}(2012)\citenamefont {Sun},
  \citenamefont {Jiang}, \citenamefont {Strzalka}, \citenamefont {Ocola},\ and\
  \citenamefont {Wang}}]{sun2012three}%
  \BibitemOpen
  \bibfield  {author} {\bibinfo {author} {\bibfnamefont {T.}~\bibnamefont
  {Sun}}, \bibinfo {author} {\bibfnamefont {Z.}~\bibnamefont {Jiang}}, \bibinfo
  {author} {\bibfnamefont {J.}~\bibnamefont {Strzalka}}, \bibinfo {author}
  {\bibfnamefont {L.}~\bibnamefont {Ocola}},\ and\ \bibinfo {author}
  {\bibfnamefont {J.}~\bibnamefont {Wang}},\ }\bibfield  {title} {\bibinfo
  {title} {Three-dimensional coherent x-ray surface scattering imaging near
  total external reflection},\ }\href@noop {} {\bibfield  {journal} {\bibinfo
  {journal} {Nature Photonics}\ }\textbf {\bibinfo {volume} {6}},\ \bibinfo
  {pages} {586} (\bibinfo {year} {2012})}\BibitemShut {NoStop}%
\bibitem [{\citenamefont {Tolan}(1999)}]{tolan1999x}%
  \BibitemOpen
  \bibfield  {author} {\bibinfo {author} {\bibfnamefont {M.}~\bibnamefont
  {Tolan}},\ }\href@noop {} {\emph {\bibinfo {title} {X-ray scattering from
  soft-matter thin films: materials science and basic research}}},\ Vol.\
  \bibinfo {volume} {148}\ (\bibinfo  {publisher} {Springer},\ \bibinfo
  {address} {Berlin},\ \bibinfo {year} {1999})\BibitemShut {NoStop}%
\bibitem [{\citenamefont {Daillant}\ and\ \citenamefont
  {Gibaud}(2008)}]{daillant2008x}%
  \BibitemOpen
  \bibfield  {author} {\bibinfo {author} {\bibfnamefont {J.}~\bibnamefont
  {Daillant}}\ and\ \bibinfo {author} {\bibfnamefont {A.}~\bibnamefont
  {Gibaud}},\ }\href@noop {} {\emph {\bibinfo {title} {X-ray and neutron
  reflectivity: principles and applications}}},\ Vol.\ \bibinfo {volume} {770}\
  (\bibinfo  {publisher} {Springer},\ \bibinfo {address} {Berlin},\ \bibinfo
  {year} {2008})\BibitemShut {NoStop}%
\bibitem [{\citenamefont {Tanksalvala}\ \emph {et~al.}(2021)\citenamefont
  {Tanksalvala}, \citenamefont {Porter}, \citenamefont {Esashi}, \citenamefont
  {Wang}, \citenamefont {Jenkins}, \citenamefont {Zhang}, \citenamefont
  {Miley}, \citenamefont {Knobloch}, \citenamefont {McBennett}, \citenamefont
  {Horiguchi} \emph {et~al.}}]{tanksalvala2021nondestructive}%
  \BibitemOpen
  \bibfield  {author} {\bibinfo {author} {\bibfnamefont {M.}~\bibnamefont
  {Tanksalvala}}, \bibinfo {author} {\bibfnamefont {C.~L.}\ \bibnamefont
  {Porter}}, \bibinfo {author} {\bibfnamefont {Y.}~\bibnamefont {Esashi}},
  \bibinfo {author} {\bibfnamefont {B.}~\bibnamefont {Wang}}, \bibinfo {author}
  {\bibfnamefont {N.~W.}\ \bibnamefont {Jenkins}}, \bibinfo {author}
  {\bibfnamefont {Z.}~\bibnamefont {Zhang}}, \bibinfo {author} {\bibfnamefont
  {G.~P.}\ \bibnamefont {Miley}}, \bibinfo {author} {\bibfnamefont {J.~L.}\
  \bibnamefont {Knobloch}}, \bibinfo {author} {\bibfnamefont {B.}~\bibnamefont
  {McBennett}}, \bibinfo {author} {\bibfnamefont {N.}~\bibnamefont
  {Horiguchi}}, \emph {et~al.},\ }\bibfield  {title} {\bibinfo {title}
  {Nondestructive, high-resolution, chemically specific 3d nanostructure
  characterization using phase-sensitive euv imaging reflectometry},\
  }\href@noop {} {\bibfield  {journal} {\bibinfo  {journal} {Science Advances}\
  }\textbf {\bibinfo {volume} {7}},\ \bibinfo {pages} {eabd9667} (\bibinfo
  {year} {2021})}\BibitemShut {NoStop}%
\bibitem [{\citenamefont {Roy}\ \emph {et~al.}(2011)\citenamefont {Roy},
  \citenamefont {Parks}, \citenamefont {Seu}, \citenamefont {Su}, \citenamefont
  {Turner}, \citenamefont {Chao}, \citenamefont {Anderson}, \citenamefont
  {Cabrini},\ and\ \citenamefont {Kevan}}]{roy2011lensless}%
  \BibitemOpen
  \bibfield  {author} {\bibinfo {author} {\bibfnamefont {S.}~\bibnamefont
  {Roy}}, \bibinfo {author} {\bibfnamefont {D.}~\bibnamefont {Parks}}, \bibinfo
  {author} {\bibfnamefont {K.}~\bibnamefont {Seu}}, \bibinfo {author}
  {\bibfnamefont {R.}~\bibnamefont {Su}}, \bibinfo {author} {\bibfnamefont
  {J.}~\bibnamefont {Turner}}, \bibinfo {author} {\bibfnamefont
  {W.}~\bibnamefont {Chao}}, \bibinfo {author} {\bibfnamefont {E.}~\bibnamefont
  {Anderson}}, \bibinfo {author} {\bibfnamefont {S.}~\bibnamefont {Cabrini}},\
  and\ \bibinfo {author} {\bibfnamefont {S.}~\bibnamefont {Kevan}},\ }\bibfield
   {title} {\bibinfo {title} {Lensless x-ray imaging in reflection geometry},\
  }\href@noop {} {\bibfield  {journal} {\bibinfo  {journal} {Nature Photonics}\
  }\textbf {\bibinfo {volume} {5}},\ \bibinfo {pages} {243} (\bibinfo {year}
  {2011})}\BibitemShut {NoStop}%
\bibitem [{\citenamefont {Withers}\ \emph {et~al.}(2021)\citenamefont
  {Withers}, \citenamefont {Bouman}, \citenamefont {Carmignato}, \citenamefont
  {Cnudde}, \citenamefont {Grimaldi}, \citenamefont {Hagen}, \citenamefont
  {Maire}, \citenamefont {Manley}, \citenamefont {Du~Plessis},\ and\
  \citenamefont {Stock}}]{withers2021x}%
  \BibitemOpen
  \bibfield  {author} {\bibinfo {author} {\bibfnamefont {P.~J.}\ \bibnamefont
  {Withers}}, \bibinfo {author} {\bibfnamefont {C.}~\bibnamefont {Bouman}},
  \bibinfo {author} {\bibfnamefont {S.}~\bibnamefont {Carmignato}}, \bibinfo
  {author} {\bibfnamefont {V.}~\bibnamefont {Cnudde}}, \bibinfo {author}
  {\bibfnamefont {D.}~\bibnamefont {Grimaldi}}, \bibinfo {author}
  {\bibfnamefont {C.~K.}\ \bibnamefont {Hagen}}, \bibinfo {author}
  {\bibfnamefont {E.}~\bibnamefont {Maire}}, \bibinfo {author} {\bibfnamefont
  {M.}~\bibnamefont {Manley}}, \bibinfo {author} {\bibfnamefont
  {A.}~\bibnamefont {Du~Plessis}},\ and\ \bibinfo {author} {\bibfnamefont
  {S.~R.}\ \bibnamefont {Stock}},\ }\bibfield  {title} {\bibinfo {title} {X-ray
  computed tomography},\ }\href@noop {} {\bibfield  {journal} {\bibinfo
  {journal} {Nature Reviews Methods Primers}\ }\textbf {\bibinfo {volume}
  {1}},\ \bibinfo {pages} {18} (\bibinfo {year} {2021})}\BibitemShut {NoStop}%
\bibitem [{\citenamefont {Chu}\ \emph {et~al.}(2023)\citenamefont {Chu},
  \citenamefont {Jiang}, \citenamefont {Wojcik}, \citenamefont {Sun},
  \citenamefont {Sprung},\ and\ \citenamefont {Wang}}]{chu2023}%
  \BibitemOpen
  \bibfield  {author} {\bibinfo {author} {\bibfnamefont {M.}~\bibnamefont
  {Chu}}, \bibinfo {author} {\bibfnamefont {Z.}~\bibnamefont {Jiang}}, \bibinfo
  {author} {\bibfnamefont {M.}~\bibnamefont {Wojcik}}, \bibinfo {author}
  {\bibfnamefont {T.}~\bibnamefont {Sun}}, \bibinfo {author} {\bibfnamefont
  {M.}~\bibnamefont {Sprung}},\ and\ \bibinfo {author} {\bibfnamefont
  {J.}~\bibnamefont {Wang}},\ }\bibfield  {title} {\bibinfo {title} {Probing
  three-dimensional mesoscopic interfacial structures in a single view using
  multibeam x-ray coherent surface scattering and holography imaging},\
  }\href@noop {} {\bibfield  {journal} {\bibinfo  {journal} {Nature
  Communications}\ }\textbf {\bibinfo {volume} {14}},\ \bibinfo {pages} {5795}
  (\bibinfo {year} {2023})}\BibitemShut {NoStop}%
\bibitem [{\citenamefont {Robinson}(1986)}]{robinson1986crystal}%
  \BibitemOpen
  \bibfield  {author} {\bibinfo {author} {\bibfnamefont {I.~K.}\ \bibnamefont
  {Robinson}},\ }\bibfield  {title} {\bibinfo {title} {Crystal truncation rods
  and surface roughness},\ }\href@noop {} {\bibfield  {journal} {\bibinfo
  {journal} {Physical Review B}\ }\textbf {\bibinfo {volume} {33}},\ \bibinfo
  {pages} {3830} (\bibinfo {year} {1986})}\BibitemShut {NoStop}%
\bibitem [{\citenamefont {Kaganer}(2007)}]{kaganer2007crystal}%
  \BibitemOpen
  \bibfield  {author} {\bibinfo {author} {\bibfnamefont {V.~M.}\ \bibnamefont
  {Kaganer}},\ }\bibfield  {title} {\bibinfo {title} {Crystal truncation rods
  in kinematical and dynamical x-ray diffraction theories},\ }\href@noop {}
  {\bibfield  {journal} {\bibinfo  {journal} {Physical Review B}\ }\textbf
  {\bibinfo {volume} {75}},\ \bibinfo {pages} {245425} (\bibinfo {year}
  {2007})}\BibitemShut {NoStop}%
\bibitem [{\citenamefont {Parratt}(1954)}]{parratt1954surface}%
  \BibitemOpen
  \bibfield  {author} {\bibinfo {author} {\bibfnamefont {L.~G.}\ \bibnamefont
  {Parratt}},\ }\bibfield  {title} {\bibinfo {title} {Surface studies of solids
  by total reflection of x-rays},\ }\href@noop {} {\bibfield  {journal}
  {\bibinfo  {journal} {Physical review}\ }\textbf {\bibinfo {volume} {95}},\
  \bibinfo {pages} {359} (\bibinfo {year} {1954})}\BibitemShut {NoStop}%
\bibitem [{\citenamefont {Myint}\ \emph {et~al.}(2023)\citenamefont {Myint},
  \citenamefont {Chu}, \citenamefont {Tripathi}, \citenamefont {Wojcik},
  \citenamefont {Zhou}, \citenamefont {Cherukara}, \citenamefont {Narayanan},
  \citenamefont {Wang},\ and\ \citenamefont {Jiang}}]{myint2023multislice}%
  \BibitemOpen
  \bibfield  {author} {\bibinfo {author} {\bibfnamefont {P.}~\bibnamefont
  {Myint}}, \bibinfo {author} {\bibfnamefont {M.}~\bibnamefont {Chu}}, \bibinfo
  {author} {\bibfnamefont {A.}~\bibnamefont {Tripathi}}, \bibinfo {author}
  {\bibfnamefont {M.~J.}\ \bibnamefont {Wojcik}}, \bibinfo {author}
  {\bibfnamefont {J.}~\bibnamefont {Zhou}}, \bibinfo {author} {\bibfnamefont
  {M.~J.}\ \bibnamefont {Cherukara}}, \bibinfo {author} {\bibfnamefont
  {S.}~\bibnamefont {Narayanan}}, \bibinfo {author} {\bibfnamefont
  {J.}~\bibnamefont {Wang}},\ and\ \bibinfo {author} {\bibfnamefont
  {Z.}~\bibnamefont {Jiang}},\ }\bibfield  {title} {\bibinfo {title}
  {Multislice forward modeling of coherent surface scattering imaging on
  surface and interfacial structures},\ }\href@noop {} {\bibfield  {journal}
  {\bibinfo  {journal} {Optics Express}\ }\textbf {\bibinfo {volume} {31}},\
  \bibinfo {pages} {11261} (\bibinfo {year} {2023})}\BibitemShut {NoStop}%
\bibitem [{\citenamefont {Li}\ \emph {et~al.}(2017)\citenamefont {Li},
  \citenamefont {Wojcik},\ and\ \citenamefont {Jacobsen}}]{li2017multislice}%
  \BibitemOpen
  \bibfield  {author} {\bibinfo {author} {\bibfnamefont {K.}~\bibnamefont
  {Li}}, \bibinfo {author} {\bibfnamefont {M.}~\bibnamefont {Wojcik}},\ and\
  \bibinfo {author} {\bibfnamefont {C.}~\bibnamefont {Jacobsen}},\ }\bibfield
  {title} {\bibinfo {title} {Multislice does it all—calculating the
  performance of nanofocusing x-ray optics},\ }\href@noop {} {\bibfield
  {journal} {\bibinfo  {journal} {Optics Express}\ }\textbf {\bibinfo {volume}
  {25}},\ \bibinfo {pages} {1831} (\bibinfo {year} {2017})}\BibitemShut
  {NoStop}%
\bibitem [{\citenamefont {Nevot}\ and\ \citenamefont
  {Croce}(1980)}]{nevot1980caracterisation}%
  \BibitemOpen
  \bibfield  {author} {\bibinfo {author} {\bibfnamefont {L.}~\bibnamefont
  {Nevot}}\ and\ \bibinfo {author} {\bibfnamefont {P.}~\bibnamefont {Croce}},\
  }\bibfield  {title} {\bibinfo {title} {Caract{\'e}risation des surfaces par
  r{\'e}flexion rasante de rayons x. application {\`a} l'{\'e}tude du polissage
  de quelques verres silicates},\ }\href@noop {} {\bibfield  {journal}
  {\bibinfo  {journal} {Revue de Physique appliqu{\'e}e}\ }\textbf {\bibinfo
  {volume} {15}},\ \bibinfo {pages} {761} (\bibinfo {year} {1980})}\BibitemShut
  {NoStop}%
\bibitem [{\citenamefont {Becker}\ \emph {et~al.}(1983)\citenamefont {Becker},
  \citenamefont {Golovchenko},\ and\ \citenamefont {Patel}}]{becker1983x}%
  \BibitemOpen
  \bibfield  {author} {\bibinfo {author} {\bibfnamefont {R.}~\bibnamefont
  {Becker}}, \bibinfo {author} {\bibfnamefont {J.~A.}\ \bibnamefont
  {Golovchenko}},\ and\ \bibinfo {author} {\bibfnamefont {J.}~\bibnamefont
  {Patel}},\ }\bibfield  {title} {\bibinfo {title} {X-ray evanescent-wave
  absorption and emission},\ }\href@noop {} {\bibfield  {journal} {\bibinfo
  {journal} {Physical review letters}\ }\textbf {\bibinfo {volume} {50}},\
  \bibinfo {pages} {153} (\bibinfo {year} {1983})}\BibitemShut {NoStop}%
\bibitem [{\citenamefont {Thibault}\ and\ \citenamefont
  {Menzel}(2013)}]{thibault2013reconstructing}%
  \BibitemOpen
  \bibfield  {author} {\bibinfo {author} {\bibfnamefont {P.}~\bibnamefont
  {Thibault}}\ and\ \bibinfo {author} {\bibfnamefont {A.}~\bibnamefont
  {Menzel}},\ }\bibfield  {title} {\bibinfo {title} {Reconstructing state
  mixtures from diffraction measurements},\ }\href@noop {} {\bibfield
  {journal} {\bibinfo  {journal} {Nature}\ }\textbf {\bibinfo {volume} {494}},\
  \bibinfo {pages} {68} (\bibinfo {year} {2013})}\BibitemShut {NoStop}%
\bibitem [{\citenamefont {Steinbrener}\ \emph {et~al.}(2010)\citenamefont
  {Steinbrener}, \citenamefont {Nelson}, \citenamefont {Huang}, \citenamefont
  {Marchesini}, \citenamefont {Shapiro}, \citenamefont {Turner},\ and\
  \citenamefont {Jacobsen}}]{steinbrener2010data}%
  \BibitemOpen
  \bibfield  {author} {\bibinfo {author} {\bibfnamefont {J.}~\bibnamefont
  {Steinbrener}}, \bibinfo {author} {\bibfnamefont {J.}~\bibnamefont {Nelson}},
  \bibinfo {author} {\bibfnamefont {X.}~\bibnamefont {Huang}}, \bibinfo
  {author} {\bibfnamefont {S.}~\bibnamefont {Marchesini}}, \bibinfo {author}
  {\bibfnamefont {D.}~\bibnamefont {Shapiro}}, \bibinfo {author} {\bibfnamefont
  {J.~J.}\ \bibnamefont {Turner}},\ and\ \bibinfo {author} {\bibfnamefont
  {C.}~\bibnamefont {Jacobsen}},\ }\bibfield  {title} {\bibinfo {title} {Data
  preparation and evaluation techniques for x-ray diffraction microscopy},\
  }\href@noop {} {\bibfield  {journal} {\bibinfo  {journal} {Optics express}\
  }\textbf {\bibinfo {volume} {18}},\ \bibinfo {pages} {18598} (\bibinfo {year}
  {2010})}\BibitemShut {NoStop}%
\bibitem [{\citenamefont {Van~Heel}\ and\ \citenamefont
  {Schatz}(2005)}]{van2005fourier}%
  \BibitemOpen
  \bibfield  {author} {\bibinfo {author} {\bibfnamefont {M.}~\bibnamefont
  {Van~Heel}}\ and\ \bibinfo {author} {\bibfnamefont {M.}~\bibnamefont
  {Schatz}},\ }\bibfield  {title} {\bibinfo {title} {Fourier shell correlation
  threshold criteria},\ }\href@noop {} {\bibfield  {journal} {\bibinfo
  {journal} {Journal of structural biology}\ }\textbf {\bibinfo {volume}
  {151}},\ \bibinfo {pages} {250} (\bibinfo {year} {2005})}\BibitemShut
  {NoStop}%
\bibitem [{\citenamefont {Vila-Comamala}\ \emph {et~al.}(2011)\citenamefont
  {Vila-Comamala}, \citenamefont {Diaz}, \citenamefont {Guizar-Sicairos},
  \citenamefont {Mantion}, \citenamefont {Kewish}, \citenamefont {Menzel},
  \citenamefont {Bunk},\ and\ \citenamefont
  {David}}]{vila2011characterization}%
  \BibitemOpen
  \bibfield  {author} {\bibinfo {author} {\bibfnamefont {J.}~\bibnamefont
  {Vila-Comamala}}, \bibinfo {author} {\bibfnamefont {A.}~\bibnamefont {Diaz}},
  \bibinfo {author} {\bibfnamefont {M.}~\bibnamefont {Guizar-Sicairos}},
  \bibinfo {author} {\bibfnamefont {A.}~\bibnamefont {Mantion}}, \bibinfo
  {author} {\bibfnamefont {C.~M.}\ \bibnamefont {Kewish}}, \bibinfo {author}
  {\bibfnamefont {A.}~\bibnamefont {Menzel}}, \bibinfo {author} {\bibfnamefont
  {O.}~\bibnamefont {Bunk}},\ and\ \bibinfo {author} {\bibfnamefont
  {C.}~\bibnamefont {David}},\ }\bibfield  {title} {\bibinfo {title}
  {Characterization of high-resolution diffractive x-ray optics by
  ptychographic coherent diffractive imaging},\ }\href@noop {} {\bibfield
  {journal} {\bibinfo  {journal} {Optics express}\ }\textbf {\bibinfo {volume}
  {19}},\ \bibinfo {pages} {21333} (\bibinfo {year} {2011})}\BibitemShut
  {NoStop}%
\bibitem [{\citenamefont {Eriksson}\ \emph {et~al.}(2014)\citenamefont
  {Eriksson}, \citenamefont {Van~der Veen},\ and\ \citenamefont
  {Quitmann}}]{eriksson2014diffraction}%
  \BibitemOpen
  \bibfield  {author} {\bibinfo {author} {\bibfnamefont {M.}~\bibnamefont
  {Eriksson}}, \bibinfo {author} {\bibfnamefont {J.~F.}\ \bibnamefont {Van~der
  Veen}},\ and\ \bibinfo {author} {\bibfnamefont {C.}~\bibnamefont
  {Quitmann}},\ }\bibfield  {title} {\bibinfo {title} {Diffraction-limited
  storage rings--a window to the science of tomorrow},\ }\href@noop {}
  {\bibfield  {journal} {\bibinfo  {journal} {Journal of synchrotron
  radiation}\ }\textbf {\bibinfo {volume} {21}},\ \bibinfo {pages} {837}
  (\bibinfo {year} {2014})}\BibitemShut {NoStop}%
\bibitem [{\citenamefont {Sinha}\ \emph {et~al.}(1988)\citenamefont {Sinha},
  \citenamefont {Sirota}, \citenamefont {Garoff},\ and\ \citenamefont
  {Stanley}}]{sinha1988x}%
  \BibitemOpen
  \bibfield  {author} {\bibinfo {author} {\bibfnamefont {S.}~\bibnamefont
  {Sinha}}, \bibinfo {author} {\bibfnamefont {E.}~\bibnamefont {Sirota}},
  \bibinfo {author} {\bibfnamefont {S.}~\bibnamefont {Garoff}},\ and\ \bibinfo
  {author} {\bibfnamefont {H.}~\bibnamefont {Stanley}},\ }\bibfield  {title}
  {\bibinfo {title} {X-ray and neutron scattering from rough surfaces},\
  }\href@noop {} {\bibfield  {journal} {\bibinfo  {journal} {Physical Review
  B}\ }\textbf {\bibinfo {volume} {38}},\ \bibinfo {pages} {2297} (\bibinfo
  {year} {1988})}\BibitemShut {NoStop}%
\bibitem [{\citenamefont {Jiang}\ \emph {et~al.}(2011)\citenamefont {Jiang},
  \citenamefont {Lee}, \citenamefont {Narayanan}, \citenamefont {Wang},\ and\
  \citenamefont {Sinha}}]{jiang2011waveguide}%
  \BibitemOpen
  \bibfield  {author} {\bibinfo {author} {\bibfnamefont {Z.}~\bibnamefont
  {Jiang}}, \bibinfo {author} {\bibfnamefont {D.~R.}\ \bibnamefont {Lee}},
  \bibinfo {author} {\bibfnamefont {S.}~\bibnamefont {Narayanan}}, \bibinfo
  {author} {\bibfnamefont {J.}~\bibnamefont {Wang}},\ and\ \bibinfo {author}
  {\bibfnamefont {S.~K.}\ \bibnamefont {Sinha}},\ }\bibfield  {title} {\bibinfo
  {title} {Waveguide-enhanced grazing-incidence small-angle x-ray scattering of
  buried nanostructures in thin films},\ }\href@noop {} {\bibfield  {journal}
  {\bibinfo  {journal} {Physical Review B}\ }\textbf {\bibinfo {volume} {84}},\
  \bibinfo {pages} {075440} (\bibinfo {year} {2011})}\BibitemShut {NoStop}%
\bibitem [{\citenamefont {Renaud}\ \emph {et~al.}(2009)\citenamefont {Renaud},
  \citenamefont {Lazzari},\ and\ \citenamefont {Leroy}}]{renaud2009probing}%
  \BibitemOpen
  \bibfield  {author} {\bibinfo {author} {\bibfnamefont {G.}~\bibnamefont
  {Renaud}}, \bibinfo {author} {\bibfnamefont {R.}~\bibnamefont {Lazzari}},\
  and\ \bibinfo {author} {\bibfnamefont {F.}~\bibnamefont {Leroy}},\ }\bibfield
   {title} {\bibinfo {title} {Probing surface and interface morphology with
  grazing incidence small angle x-ray scattering},\ }\href@noop {} {\bibfield
  {journal} {\bibinfo  {journal} {Surface Science Reports}\ }\textbf {\bibinfo
  {volume} {64}},\ \bibinfo {pages} {255} (\bibinfo {year} {2009})}\BibitemShut
  {NoStop}%
\bibitem [{\citenamefont {Hexemer}\ and\ \citenamefont
  {M{\"u}ller-Buschbaum}(2015)}]{hexemer2015advanced}%
  \BibitemOpen
  \bibfield  {author} {\bibinfo {author} {\bibfnamefont {A.}~\bibnamefont
  {Hexemer}}\ and\ \bibinfo {author} {\bibfnamefont {P.}~\bibnamefont
  {M{\"u}ller-Buschbaum}},\ }\bibfield  {title} {\bibinfo {title} {Advanced
  grazing-incidence techniques for modern soft-matter materials analysis},\
  }\href@noop {} {\bibfield  {journal} {\bibinfo  {journal} {IUCrJ}\ }\textbf
  {\bibinfo {volume} {2}},\ \bibinfo {pages} {106} (\bibinfo {year}
  {2015})}\BibitemShut {NoStop}%
\bibitem [{\citenamefont {Li}\ \emph {et~al.}(2016)\citenamefont {Li},
  \citenamefont {Senesi},\ and\ \citenamefont {Lee}}]{li2016small}%
  \BibitemOpen
  \bibfield  {author} {\bibinfo {author} {\bibfnamefont {T.}~\bibnamefont
  {Li}}, \bibinfo {author} {\bibfnamefont {A.~J.}\ \bibnamefont {Senesi}},\
  and\ \bibinfo {author} {\bibfnamefont {B.}~\bibnamefont {Lee}},\ }\bibfield
  {title} {\bibinfo {title} {Small angle x-ray scattering for nanoparticle
  research},\ }\href@noop {} {\bibfield  {journal} {\bibinfo  {journal}
  {Chemical reviews}\ }\textbf {\bibinfo {volume} {116}},\ \bibinfo {pages}
  {11128} (\bibinfo {year} {2016})}\BibitemShut {NoStop}%
\bibitem [{\citenamefont {Yang}\ and\ \citenamefont
  {Sinha}(2023)}]{yang2023three}%
  \BibitemOpen
  \bibfield  {author} {\bibinfo {author} {\bibfnamefont {Y.}~\bibnamefont
  {Yang}}\ and\ \bibinfo {author} {\bibfnamefont {S.~K.}\ \bibnamefont
  {Sinha}},\ }\bibfield  {title} {\bibinfo {title} {Three-dimensional imaging
  using coherent x rays at grazing incidence geometry},\ }\href@noop {}
  {\bibfield  {journal} {\bibinfo  {journal} {JOSA A}\ }\textbf {\bibinfo
  {volume} {40}},\ \bibinfo {pages} {1500} (\bibinfo {year}
  {2023})}\BibitemShut {NoStop}%
\bibitem [{\citenamefont {Pfeiffer}\ \emph {et~al.}(2002)\citenamefont
  {Pfeiffer}, \citenamefont {David}, \citenamefont {Burghammer}, \citenamefont
  {Riekel},\ and\ \citenamefont {Salditt}}]{pfeiffer2002two}%
  \BibitemOpen
  \bibfield  {author} {\bibinfo {author} {\bibfnamefont {F.}~\bibnamefont
  {Pfeiffer}}, \bibinfo {author} {\bibfnamefont {C.}~\bibnamefont {David}},
  \bibinfo {author} {\bibfnamefont {M.}~\bibnamefont {Burghammer}}, \bibinfo
  {author} {\bibfnamefont {C.}~\bibnamefont {Riekel}},\ and\ \bibinfo {author}
  {\bibfnamefont {T.}~\bibnamefont {Salditt}},\ }\bibfield  {title} {\bibinfo
  {title} {Two-dimensional x-ray waveguides and point sources},\ }\href@noop {}
  {\bibfield  {journal} {\bibinfo  {journal} {Science}\ }\textbf {\bibinfo
  {volume} {297}},\ \bibinfo {pages} {230} (\bibinfo {year}
  {2002})}\BibitemShut {NoStop}%
\bibitem [{\citenamefont {Zegenhagen}\ and\ \citenamefont
  {Kazimirov}(2013)}]{zegenhagen2013x}%
  \BibitemOpen
  \bibfield  {author} {\bibinfo {author} {\bibfnamefont {J.}~\bibnamefont
  {Zegenhagen}}\ and\ \bibinfo {author} {\bibfnamefont {A.}~\bibnamefont
  {Kazimirov}},\ }\href@noop {} {\emph {\bibinfo {title} {X-ray Standing Wave
  Technique, The: Principles And Applications}}},\ Vol.~\bibinfo {volume} {7}\
  (\bibinfo  {publisher} {World Scientific},\ \bibinfo {address} {Singapore},\
  \bibinfo {year} {2013})\BibitemShut {NoStop}%
\bibitem [{\citenamefont {Wang}\ \emph {et~al.}(1992)\citenamefont {Wang},
  \citenamefont {Bedzyk},\ and\ \citenamefont {Caffrey}}]{wang1992resonance}%
  \BibitemOpen
  \bibfield  {author} {\bibinfo {author} {\bibfnamefont {J.}~\bibnamefont
  {Wang}}, \bibinfo {author} {\bibfnamefont {M.~J.}\ \bibnamefont {Bedzyk}},\
  and\ \bibinfo {author} {\bibfnamefont {M.}~\bibnamefont {Caffrey}},\
  }\bibfield  {title} {\bibinfo {title} {Resonance-enhanced x-rays in thin
  films: a structure probe for membranes and surface layers},\ }\href@noop {}
  {\bibfield  {journal} {\bibinfo  {journal} {Science}\ }\textbf {\bibinfo
  {volume} {258}},\ \bibinfo {pages} {775} (\bibinfo {year}
  {1992})}\BibitemShut {NoStop}%
\bibitem [{\citenamefont {Maiden}\ \emph {et~al.}(2012)\citenamefont {Maiden},
  \citenamefont {Humphry},\ and\ \citenamefont
  {Rodenburg}}]{maiden2012ptychographic}%
  \BibitemOpen
  \bibfield  {author} {\bibinfo {author} {\bibfnamefont {A.~M.}\ \bibnamefont
  {Maiden}}, \bibinfo {author} {\bibfnamefont {M.~J.}\ \bibnamefont
  {Humphry}},\ and\ \bibinfo {author} {\bibfnamefont {J.~M.}\ \bibnamefont
  {Rodenburg}},\ }\bibfield  {title} {\bibinfo {title} {Ptychographic
  transmission microscopy in three dimensions using a multi-slice approach},\
  }\href@noop {} {\bibfield  {journal} {\bibinfo  {journal} {JOSA A}\ }\textbf
  {\bibinfo {volume} {29}},\ \bibinfo {pages} {1606} (\bibinfo {year}
  {2012})}\BibitemShut {NoStop}%
\bibitem [{\citenamefont {Shimomura}\ \emph {et~al.}(2015)\citenamefont
  {Shimomura}, \citenamefont {Suzuki}, \citenamefont {Hirose},\ and\
  \citenamefont {Takahashi}}]{shimomura2015precession}%
  \BibitemOpen
  \bibfield  {author} {\bibinfo {author} {\bibfnamefont {K.}~\bibnamefont
  {Shimomura}}, \bibinfo {author} {\bibfnamefont {A.}~\bibnamefont {Suzuki}},
  \bibinfo {author} {\bibfnamefont {M.}~\bibnamefont {Hirose}},\ and\ \bibinfo
  {author} {\bibfnamefont {Y.}~\bibnamefont {Takahashi}},\ }\bibfield  {title}
  {\bibinfo {title} {Precession x-ray ptychography with multislice approach},\
  }\href@noop {} {\bibfield  {journal} {\bibinfo  {journal} {Physical Review
  B}\ }\textbf {\bibinfo {volume} {91}},\ \bibinfo {pages} {214114} (\bibinfo
  {year} {2015})}\BibitemShut {NoStop}%
\bibitem [{\citenamefont {Tsai}\ \emph {et~al.}(2016)\citenamefont {Tsai},
  \citenamefont {Usov}, \citenamefont {Diaz}, \citenamefont {Menzel},\ and\
  \citenamefont {Guizar-Sicairos}}]{tsai2016x}%
  \BibitemOpen
  \bibfield  {author} {\bibinfo {author} {\bibfnamefont {E.~H.}\ \bibnamefont
  {Tsai}}, \bibinfo {author} {\bibfnamefont {I.}~\bibnamefont {Usov}}, \bibinfo
  {author} {\bibfnamefont {A.}~\bibnamefont {Diaz}}, \bibinfo {author}
  {\bibfnamefont {A.}~\bibnamefont {Menzel}},\ and\ \bibinfo {author}
  {\bibfnamefont {M.}~\bibnamefont {Guizar-Sicairos}},\ }\bibfield  {title}
  {\bibinfo {title} {X-ray ptychography with extended depth of field},\
  }\href@noop {} {\bibfield  {journal} {\bibinfo  {journal} {Optics express}\
  }\textbf {\bibinfo {volume} {24}},\ \bibinfo {pages} {29089} (\bibinfo {year}
  {2016})}\BibitemShut {NoStop}%
\bibitem [{\citenamefont {{\"O}zt{\"u}rk}\ \emph {et~al.}(2018)\citenamefont
  {{\"O}zt{\"u}rk}, \citenamefont {Yan}, \citenamefont {He}, \citenamefont
  {Ge}, \citenamefont {Dong}, \citenamefont {Lin}, \citenamefont {Nazaretski},
  \citenamefont {Robinson}, \citenamefont {Chu},\ and\ \citenamefont
  {Huang}}]{ozturk2018multi}%
  \BibitemOpen
  \bibfield  {author} {\bibinfo {author} {\bibfnamefont {H.}~\bibnamefont
  {{\"O}zt{\"u}rk}}, \bibinfo {author} {\bibfnamefont {H.}~\bibnamefont {Yan}},
  \bibinfo {author} {\bibfnamefont {Y.}~\bibnamefont {He}}, \bibinfo {author}
  {\bibfnamefont {M.}~\bibnamefont {Ge}}, \bibinfo {author} {\bibfnamefont
  {Z.}~\bibnamefont {Dong}}, \bibinfo {author} {\bibfnamefont {M.}~\bibnamefont
  {Lin}}, \bibinfo {author} {\bibfnamefont {E.}~\bibnamefont {Nazaretski}},
  \bibinfo {author} {\bibfnamefont {I.~K.}\ \bibnamefont {Robinson}}, \bibinfo
  {author} {\bibfnamefont {Y.~S.}\ \bibnamefont {Chu}},\ and\ \bibinfo {author}
  {\bibfnamefont {X.}~\bibnamefont {Huang}},\ }\bibfield  {title} {\bibinfo
  {title} {Multi-slice ptychography with large numerical aperture multilayer
  laue lenses},\ }\href@noop {} {\bibfield  {journal} {\bibinfo  {journal}
  {Optica}\ }\textbf {\bibinfo {volume} {5}},\ \bibinfo {pages} {601} (\bibinfo
  {year} {2018})}\BibitemShut {NoStop}%
\bibitem [{\citenamefont {Sasaki}\ \emph {et~al.}(1994)\citenamefont {Sasaki},
  \citenamefont {Suzuki},\ and\ \citenamefont
  {Ishibashi}}]{sasaki1994fluorescent}%
  \BibitemOpen
  \bibfield  {author} {\bibinfo {author} {\bibfnamefont {Y.}~\bibnamefont
  {Sasaki}}, \bibinfo {author} {\bibfnamefont {Y.}~\bibnamefont {Suzuki}},\
  and\ \bibinfo {author} {\bibfnamefont {T.}~\bibnamefont {Ishibashi}},\
  }\bibfield  {title} {\bibinfo {title} {Fluorescent x-ray interference from a
  protein monolayer},\ }\href@noop {} {\bibfield  {journal} {\bibinfo
  {journal} {Science}\ }\textbf {\bibinfo {volume} {263}},\ \bibinfo {pages}
  {62} (\bibinfo {year} {1994})}\BibitemShut {NoStop}%
\bibitem [{\citenamefont {Jiang}\ \emph {et~al.}(2020)\citenamefont {Jiang},
  \citenamefont {Strzalka}, \citenamefont {Walko},\ and\ \citenamefont
  {Wang}}]{jiang2020reconstruction}%
  \BibitemOpen
  \bibfield  {author} {\bibinfo {author} {\bibfnamefont {Z.}~\bibnamefont
  {Jiang}}, \bibinfo {author} {\bibfnamefont {J.~W.}\ \bibnamefont {Strzalka}},
  \bibinfo {author} {\bibfnamefont {D.~A.}\ \bibnamefont {Walko}},\ and\
  \bibinfo {author} {\bibfnamefont {J.}~\bibnamefont {Wang}},\ }\bibfield
  {title} {\bibinfo {title} {Reconstruction of evolving nanostructures in
  ultrathin films with x-ray waveguide fluorescence holography},\ }\href@noop
  {} {\bibfield  {journal} {\bibinfo  {journal} {Nature Communications}\
  }\textbf {\bibinfo {volume} {11}},\ \bibinfo {pages} {3197} (\bibinfo {year}
  {2020})}\BibitemShut {NoStop}%
\bibitem [{\citenamefont {Zhu}\ \emph {et~al.}(2015)\citenamefont {Zhu},
  \citenamefont {Harder}, \citenamefont {Diaz}, \citenamefont {Komanicky},
  \citenamefont {Barbour}, \citenamefont {Xu}, \citenamefont {Huang},
  \citenamefont {Liu}, \citenamefont {Pierce}, \citenamefont {Menzel} \emph
  {et~al.}}]{zhu2015ptychographic}%
  \BibitemOpen
  \bibfield  {author} {\bibinfo {author} {\bibfnamefont {C.}~\bibnamefont
  {Zhu}}, \bibinfo {author} {\bibfnamefont {R.}~\bibnamefont {Harder}},
  \bibinfo {author} {\bibfnamefont {A.}~\bibnamefont {Diaz}}, \bibinfo {author}
  {\bibfnamefont {V.}~\bibnamefont {Komanicky}}, \bibinfo {author}
  {\bibfnamefont {A.}~\bibnamefont {Barbour}}, \bibinfo {author} {\bibfnamefont
  {R.}~\bibnamefont {Xu}}, \bibinfo {author} {\bibfnamefont {X.}~\bibnamefont
  {Huang}}, \bibinfo {author} {\bibfnamefont {Y.}~\bibnamefont {Liu}}, \bibinfo
  {author} {\bibfnamefont {M.~S.}\ \bibnamefont {Pierce}}, \bibinfo {author}
  {\bibfnamefont {A.}~\bibnamefont {Menzel}}, \emph {et~al.},\ }\bibfield
  {title} {\bibinfo {title} {Ptychographic x-ray imaging of surfaces on crystal
  truncation rod},\ }\href@noop {} {\bibfield  {journal} {\bibinfo  {journal}
  {Applied Physics Letters}\ }\textbf {\bibinfo {volume} {106}} (\bibinfo
  {year} {2015})}\BibitemShut {NoStop}%
\bibitem [{\citenamefont {Huang}\ \emph {et~al.}(2014)\citenamefont {Huang},
  \citenamefont {Yan}, \citenamefont {Harder}, \citenamefont {Hwu},
  \citenamefont {Robinson},\ and\ \citenamefont {Chu}}]{huang2014optimization}%
  \BibitemOpen
  \bibfield  {author} {\bibinfo {author} {\bibfnamefont {X.}~\bibnamefont
  {Huang}}, \bibinfo {author} {\bibfnamefont {H.}~\bibnamefont {Yan}}, \bibinfo
  {author} {\bibfnamefont {R.}~\bibnamefont {Harder}}, \bibinfo {author}
  {\bibfnamefont {Y.}~\bibnamefont {Hwu}}, \bibinfo {author} {\bibfnamefont
  {I.~K.}\ \bibnamefont {Robinson}},\ and\ \bibinfo {author} {\bibfnamefont
  {Y.~S.}\ \bibnamefont {Chu}},\ }\bibfield  {title} {\bibinfo {title}
  {Optimization of overlap uniformness for ptychography},\ }\href@noop {}
  {\bibfield  {journal} {\bibinfo  {journal} {Optics Express}\ }\textbf
  {\bibinfo {volume} {22}},\ \bibinfo {pages} {12634} (\bibinfo {year}
  {2014})}\BibitemShut {NoStop}%
\bibitem [{\citenamefont {Maiden}\ \emph {et~al.}(2017)\citenamefont {Maiden},
  \citenamefont {Johnson},\ and\ \citenamefont {Li}}]{maiden2017further}%
  \BibitemOpen
  \bibfield  {author} {\bibinfo {author} {\bibfnamefont {A.}~\bibnamefont
  {Maiden}}, \bibinfo {author} {\bibfnamefont {D.}~\bibnamefont {Johnson}},\
  and\ \bibinfo {author} {\bibfnamefont {P.}~\bibnamefont {Li}},\ }\bibfield
  {title} {\bibinfo {title} {Further improvements to the ptychographical
  iterative engine},\ }\href@noop {} {\bibfield  {journal} {\bibinfo  {journal}
  {Optica}\ }\textbf {\bibinfo {volume} {4}},\ \bibinfo {pages} {736} (\bibinfo
  {year} {2017})}\BibitemShut {NoStop}%
\bibitem [{\citenamefont {Tripathi}\ \emph {et~al.}(2014)\citenamefont
  {Tripathi}, \citenamefont {McNulty},\ and\ \citenamefont
  {Shpyrko}}]{Tripathi:14}%
  \BibitemOpen
  \bibfield  {author} {\bibinfo {author} {\bibfnamefont {A.}~\bibnamefont
  {Tripathi}}, \bibinfo {author} {\bibfnamefont {I.}~\bibnamefont {McNulty}},\
  and\ \bibinfo {author} {\bibfnamefont {O.~G.}\ \bibnamefont {Shpyrko}},\
  }\bibfield  {title} {\bibinfo {title} {Ptychographic overlap constraint
  errors and the limits of their numerical recovery using conjugate gradient
  descent methods},\ }\href@noop {} {\bibfield  {journal} {\bibinfo  {journal}
  {Opt. Express}\ }\textbf {\bibinfo {volume} {22}},\ \bibinfo {pages} {1452}
  (\bibinfo {year} {2014})}\BibitemShut {NoStop}%
\bibitem [{\citenamefont {Bottou}(2010)}]{minibatch_grad_intro}%
  \BibitemOpen
  \bibfield  {author} {\bibinfo {author} {\bibfnamefont {L.}~\bibnamefont
  {Bottou}},\ }\bibfield  {title} {\bibinfo {title} {Large-scale machine
  learning with stochastic gradient descent},\ }in\ \href@noop {} {\emph
  {\bibinfo {booktitle} {Proceedings of COMPSTAT'2010}}},\ \bibinfo {editor}
  {edited by\ \bibinfo {editor} {\bibfnamefont {Y.}~\bibnamefont
  {Lechevallier}}\ and\ \bibinfo {editor} {\bibfnamefont {G.}~\bibnamefont
  {Saporta}}}\ (\bibinfo  {publisher} {Physica-Verlag HD},\ \bibinfo {address}
  {Heidelberg},\ \bibinfo {year} {2010})\ pp.\ \bibinfo {pages}
  {177--186}\BibitemShut {NoStop}%
\bibitem [{\citenamefont {Gursoy}\ and\ \citenamefont {Ching}(2022)}]{Tike}%
  \BibitemOpen
  \bibfield  {author} {\bibinfo {author} {\bibfnamefont {D.}~\bibnamefont
  {Gursoy}}\ and\ \bibinfo {author} {\bibfnamefont {D.~J.}\ \bibnamefont
  {Ching}},\ }\href@noop {} {\bibinfo {title} {Tike}},\ \bibinfo {howpublished}
  {[Computer Software] \url{https://doi.org/10.11578/dc.20230202.1}} (\bibinfo
  {year} {2022})\BibitemShut {NoStop}%
\end{thebibliography}%


\begin{thebibliography}{10}%
\makeatletter
\providecommand \@ifxundefined [1]{%
 \@ifx{#1\undefined}
}%
\providecommand \@ifnum [1]{%
 \ifnum #1\expandafter \@firstoftwo
 \else \expandafter \@secondoftwo
 \fi
}%
\providecommand \@ifx [1]{%
 \ifx #1\expandafter \@firstoftwo
 \else \expandafter \@secondoftwo
 \fi
}%
\providecommand \natexlab [1]{#1}%
\providecommand \enquote  [1]{``#1''}%
\providecommand \bibnamefont  [1]{#1}%
\providecommand \bibfnamefont [1]{#1}%
\providecommand \citenamefont [1]{#1}%
\providecommand \href@noop [0]{\@secondoftwo}%
\providecommand \href [0]{\begingroup \@sanitize@url \@href}%
\providecommand \@href[1]{\@@startlink{#1}\@@href}%
\providecommand \@@href[1]{\endgroup#1\@@endlink}%
\providecommand \@sanitize@url [0]{\catcode `\\12\catcode `\$12\catcode
  `\&12\catcode `\#12\catcode `\^12\catcode `\_12\catcode `\%12\relax}%
\providecommand \@@startlink[1]{}%
\providecommand \@@endlink[0]{}%
\providecommand \url  [0]{\begingroup\@sanitize@url \@url }%
\providecommand \@url [1]{\endgroup\@href {#1}{\urlprefix }}%
\providecommand \urlprefix  [0]{URL }%
\providecommand \Eprint [0]{\href }%
\providecommand \doibase [0]{https://doi.org/}%
\providecommand \selectlanguage [0]{\@gobble}%
\providecommand \bibinfo  [0]{\@secondoftwo}%
\providecommand \bibfield  [0]{\@secondoftwo}%
\providecommand \translation [1]{[#1]}%
\providecommand \BibitemOpen [0]{}%
\providecommand \bibitemStop [0]{}%
\providecommand \bibitemNoStop [0]{.\EOS\space}%
\providecommand \EOS [0]{\spacefactor3000\relax}%
\providecommand \BibitemShut  [1]{\csname bibitem#1\endcsname}%
\let\auto@bib@innerbib\@empty
\bibitem [{\citenamefont {Paganin}(2013)}]{paganin2013coherent}%
  \BibitemOpen
  \bibfield  {author} {\bibinfo {author} {\bibfnamefont {D.}~\bibnamefont
  {Paganin}},\ }\href@noop {} {\emph {\bibinfo {title} {Coherent X-Ray
  Optics}}},\ Oxford Series on Synchrotron Radiation\ (\bibinfo  {publisher}
  {Oxford University Press},\ \bibinfo {address} {Oxford},\ \bibinfo {year}
  {2013})\BibitemShut {NoStop}%
\bibitem [{\citenamefont {Odstr\v{c}il}\ \emph {et~al.}(2018)\citenamefont
  {Odstr\v{c}il}, \citenamefont {Menzel},\ and\ \citenamefont
  {Guizar-Sicairos}}]{Odstrcil:18}%
  \BibitemOpen
  \bibfield  {author} {\bibinfo {author} {\bibfnamefont {M.}~\bibnamefont
  {Odstr\v{c}il}}, \bibinfo {author} {\bibfnamefont {A.}~\bibnamefont
  {Menzel}},\ and\ \bibinfo {author} {\bibfnamefont {M.}~\bibnamefont
  {Guizar-Sicairos}},\ }\bibfield  {title} {\bibinfo {title} {Iterative
  least-squares solver for generalized maximum-likelihood ptychography},\
  }\href@noop {} {\bibfield  {journal} {\bibinfo  {journal} {Opt. Express}\
  }\textbf {\bibinfo {volume} {26}},\ \bibinfo {pages} {3108} (\bibinfo {year}
  {2018})}\BibitemShut {NoStop}%
\bibitem [{\citenamefont {Maiden}\ and\ \citenamefont
  {Rodenburg}(2009)}]{maiden2009improved}%
  \BibitemOpen
  \bibfield  {author} {\bibinfo {author} {\bibfnamefont {A.~M.}\ \bibnamefont
  {Maiden}}\ and\ \bibinfo {author} {\bibfnamefont {J.~M.}\ \bibnamefont
  {Rodenburg}},\ }\bibfield  {title} {\bibinfo {title} {An improved
  ptychographical phase retrieval algorithm for diffractive imaging},\
  }\href@noop {} {\bibfield  {journal} {\bibinfo  {journal} {Ultramicroscopy}\
  }\textbf {\bibinfo {volume} {109}},\ \bibinfo {pages} {1256} (\bibinfo {year}
  {2009})}\BibitemShut {NoStop}%
\bibitem [{\citenamefont {Tripathi}\ \emph {et~al.}(2014)\citenamefont
  {Tripathi}, \citenamefont {McNulty},\ and\ \citenamefont
  {Shpyrko}}]{Tripathi:14}%
  \BibitemOpen
  \bibfield  {author} {\bibinfo {author} {\bibfnamefont {A.}~\bibnamefont
  {Tripathi}}, \bibinfo {author} {\bibfnamefont {I.}~\bibnamefont {McNulty}},\
  and\ \bibinfo {author} {\bibfnamefont {O.~G.}\ \bibnamefont {Shpyrko}},\
  }\bibfield  {title} {\bibinfo {title} {Ptychographic overlap constraint
  errors and the limits of their numerical recovery using conjugate gradient
  descent methods},\ }\href@noop {} {\bibfield  {journal} {\bibinfo  {journal}
  {Opt. Express}\ }\textbf {\bibinfo {volume} {22}},\ \bibinfo {pages} {1452}
  (\bibinfo {year} {2014})}\BibitemShut {NoStop}%
\bibitem [{\citenamefont {Chu}\ \emph {et~al.}(2023)\citenamefont {Chu},
  \citenamefont {Jiang}, \citenamefont {Wojcik}, \citenamefont {Sun},
  \citenamefont {Sprung},\ and\ \citenamefont {Wang}}]{chu2023}%
  \BibitemOpen
  \bibfield  {author} {\bibinfo {author} {\bibfnamefont {M.}~\bibnamefont
  {Chu}}, \bibinfo {author} {\bibfnamefont {Z.}~\bibnamefont {Jiang}}, \bibinfo
  {author} {\bibfnamefont {M.}~\bibnamefont {Wojcik}}, \bibinfo {author}
  {\bibfnamefont {T.}~\bibnamefont {Sun}}, \bibinfo {author} {\bibfnamefont
  {M.}~\bibnamefont {Sprung}},\ and\ \bibinfo {author} {\bibfnamefont
  {J.}~\bibnamefont {Wang}},\ }\bibfield  {title} {\bibinfo {title} {Probing
  three-dimensional mesoscopic interfacial structures in a single view using
  multibeam x-ray coherent surface scattering and holography imaging},\
  }\href@noop {} {\bibfield  {journal} {\bibinfo  {journal} {Nature
  Communications}\ }\textbf {\bibinfo {volume} {14}},\ \bibinfo {pages} {5795}
  (\bibinfo {year} {2023})}\BibitemShut {NoStop}%
\bibitem [{\citenamefont {Myint}\ \emph {et~al.}(2023)\citenamefont {Myint},
  \citenamefont {Chu}, \citenamefont {Tripathi}, \citenamefont {Wojcik},
  \citenamefont {Zhou}, \citenamefont {Cherukara}, \citenamefont {Narayanan},
  \citenamefont {Wang},\ and\ \citenamefont {Jiang}}]{myint2023multislice}%
  \BibitemOpen
  \bibfield  {author} {\bibinfo {author} {\bibfnamefont {P.}~\bibnamefont
  {Myint}}, \bibinfo {author} {\bibfnamefont {M.}~\bibnamefont {Chu}}, \bibinfo
  {author} {\bibfnamefont {A.}~\bibnamefont {Tripathi}}, \bibinfo {author}
  {\bibfnamefont {M.~J.}\ \bibnamefont {Wojcik}}, \bibinfo {author}
  {\bibfnamefont {J.}~\bibnamefont {Zhou}}, \bibinfo {author} {\bibfnamefont
  {M.~J.}\ \bibnamefont {Cherukara}}, \bibinfo {author} {\bibfnamefont
  {S.}~\bibnamefont {Narayanan}}, \bibinfo {author} {\bibfnamefont
  {J.}~\bibnamefont {Wang}},\ and\ \bibinfo {author} {\bibfnamefont
  {Z.}~\bibnamefont {Jiang}},\ }\bibfield  {title} {\bibinfo {title}
  {Multislice forward modeling of coherent surface scattering imaging on
  surface and interfacial structures},\ }\href@noop {} {\bibfield  {journal}
  {\bibinfo  {journal} {Optics Express}\ }\textbf {\bibinfo {volume} {31}},\
  \bibinfo {pages} {11261} (\bibinfo {year} {2023})}\BibitemShut {NoStop}%
\bibitem [{\citenamefont {Huang}\ \emph {et~al.}(2014)\citenamefont {Huang},
  \citenamefont {Yan}, \citenamefont {Harder}, \citenamefont {Hwu},
  \citenamefont {Robinson},\ and\ \citenamefont {Chu}}]{huang2014optimization}%
  \BibitemOpen
  \bibfield  {author} {\bibinfo {author} {\bibfnamefont {X.}~\bibnamefont
  {Huang}}, \bibinfo {author} {\bibfnamefont {H.}~\bibnamefont {Yan}}, \bibinfo
  {author} {\bibfnamefont {R.}~\bibnamefont {Harder}}, \bibinfo {author}
  {\bibfnamefont {Y.}~\bibnamefont {Hwu}}, \bibinfo {author} {\bibfnamefont
  {I.~K.}\ \bibnamefont {Robinson}},\ and\ \bibinfo {author} {\bibfnamefont
  {Y.~S.}\ \bibnamefont {Chu}},\ }\bibfield  {title} {\bibinfo {title}
  {Optimization of overlap uniformness for ptychography},\ }\href@noop {}
  {\bibfield  {journal} {\bibinfo  {journal} {Optics Express}\ }\textbf
  {\bibinfo {volume} {22}},\ \bibinfo {pages} {12634} (\bibinfo {year}
  {2014})}\BibitemShut {NoStop}%
\bibitem [{\citenamefont {Sinha}\ \emph {et~al.}(1988)\citenamefont {Sinha},
  \citenamefont {Sirota}, \citenamefont {Garoff},\ and\ \citenamefont
  {Stanley}}]{sinha1988x}%
  \BibitemOpen
  \bibfield  {author} {\bibinfo {author} {\bibfnamefont {S.}~\bibnamefont
  {Sinha}}, \bibinfo {author} {\bibfnamefont {E.}~\bibnamefont {Sirota}},
  \bibinfo {author} {\bibfnamefont {S.}~\bibnamefont {Garoff}},\ and\ \bibinfo
  {author} {\bibfnamefont {H.}~\bibnamefont {Stanley}},\ }\bibfield  {title}
  {\bibinfo {title} {X-ray and neutron scattering from rough surfaces},\
  }\href@noop {} {\bibfield  {journal} {\bibinfo  {journal} {Physical Review
  B}\ }\textbf {\bibinfo {volume} {38}},\ \bibinfo {pages} {2297} (\bibinfo
  {year} {1988})}\BibitemShut {NoStop}%
\bibitem [{\citenamefont {Renaud}\ \emph {et~al.}(2009)\citenamefont {Renaud},
  \citenamefont {Lazzari},\ and\ \citenamefont {Leroy}}]{renaud2009probing}%
  \BibitemOpen
  \bibfield  {author} {\bibinfo {author} {\bibfnamefont {G.}~\bibnamefont
  {Renaud}}, \bibinfo {author} {\bibfnamefont {R.}~\bibnamefont {Lazzari}},\
  and\ \bibinfo {author} {\bibfnamefont {F.}~\bibnamefont {Leroy}},\ }\bibfield
   {title} {\bibinfo {title} {Probing surface and interface morphology with
  grazing incidence small angle x-ray scattering},\ }\href@noop {} {\bibfield
  {journal} {\bibinfo  {journal} {Surface Science Reports}\ }\textbf {\bibinfo
  {volume} {64}},\ \bibinfo {pages} {255} (\bibinfo {year} {2009})}\BibitemShut
  {NoStop}%
\bibitem [{\citenamefont {Jiang}\ \emph {et~al.}(2011)\citenamefont {Jiang},
  \citenamefont {Lee}, \citenamefont {Narayanan}, \citenamefont {Wang},\ and\
  \citenamefont {Sinha}}]{jiang2011waveguide}%
  \BibitemOpen
  \bibfield  {author} {\bibinfo {author} {\bibfnamefont {Z.}~\bibnamefont
  {Jiang}}, \bibinfo {author} {\bibfnamefont {D.~R.}\ \bibnamefont {Lee}},
  \bibinfo {author} {\bibfnamefont {S.}~\bibnamefont {Narayanan}}, \bibinfo
  {author} {\bibfnamefont {J.}~\bibnamefont {Wang}},\ and\ \bibinfo {author}
  {\bibfnamefont {S.~K.}\ \bibnamefont {Sinha}},\ }\bibfield  {title} {\bibinfo
  {title} {Waveguide-enhanced grazing-incidence small-angle x-ray scattering of
  buried nanostructures in thin films},\ }\href@noop {} {\bibfield  {journal}
  {\bibinfo  {journal} {Physical Review B}\ }\textbf {\bibinfo {volume} {84}},\
  \bibinfo {pages} {075440} (\bibinfo {year} {2011})}\BibitemShut {NoStop}%
\end{thebibliography}%

\end{document}



\title[Article Title]{\centering Supplementary Materials:\\ Three-dimensional Hard X-ray Ptychographic Reflectometry Imaging on Extended Mesoscopic Surface Structures}

\maketitle

\tableofcontents

\newpage
\section{Method for ptychography reconstruction} \label{supp:ptycho}
\noindent
Define the measurement array sizes to be $M \times N$ and the sample array sizes to be $ M_T \times N_T$, with $M \ll M_T$ and $N \ll N_T$, where $\{ M, N, M_T, N_T \} \in \mathds{Z}^+$.
Next, let $s \in \{ 1, 2, \cdots, N_s \}$ define the scan position (and thus measurement) indices for the total $N_s$ scan positions, and let $p \in \{ 1, 2, \cdots, N_p \}$ be the Spatial Coherence Probe Mode (SCPM) indices for the total $N_p$ modes assumed.
We define the local sample view at scan position $s$ to be (in vector notation) $T_s \in \mathds{C}^{ M N \times 1}$ and the total sample field of view (FOV) to be $T \in \mathds{C}^{ M_T N_T \times 1}$. We define the SCPMs to be $\phi_p \in \mathds{C}^{ M N \times 1}$, with additional SCPM notation as $\Phi_p \in \mathds{C}^{ M_T N_T \times 1}$ defined as the original SCPMs in array notation $\text{mat}\left( \phi_p \right) \in \mathds{C}^{M \times N}$ zero padded both horizontally and vertically to be of the same size as the total sample array $\text{mat}\left(T\right) \in \mathds{C}^{M_T \times N_T}$, and then reshaped to be a vector of size $M_T N_T \times 1$. Furthermore, we define the notation $\Phi_{sp}$ to mean that this SCPM is shifted to the scan position $s$ within the total sample array. 

\subsection{Traditional PIE} \label{supp:traditionalPIE}
\noindent
The exit wave under the projection approximation \cite{paganin2013coherent} at scan position $s$ and for SCPM index $p$ is given by 
\begin{align} \label{supp:Eqn:proj_approx_lower}
\psi_{sp} = \text{diag}\left( \phi_p \right) T_s\in \mathds{C}^{M N \times 1},
\end{align}
where the $\text{diag}(\mathbf{a})$ notation means constructing a matrix of all zeros except its diagonal which contains the elements of the vector $\mathbf{a}$. We first enforce the diffraction measurement constraint on these exitwaves using either a Gaussian or Poisson noise model \cite{Odstrcil:18} to arrive at the optimized exit wave $\psi^{\text{opt}}_{sp}\in \mathds{C}^{M N \times 1}$. 
From this, the traditional Ptychographic Iterative Engine (PIE) \cite{maiden2009improved} update rules are given by:
\begin{align}  
 T^{(k+1)}_{s} &= \label{supp:Eqn:seq_rPIE_update_T}
T_s^{(k)} + 
\\
&\quad  \nonumber
\bigg[ \sum_{p} 
\text{diag}\Big( \big\vert \phi^{(k)}_{p} \big\vert^2 
+
v^{(k)}_{\sigma}  
\Big)
\bigg]^{-1}
\bigg[
\sum_{p} 
 \text{diag}\big( \phi^{(k)}_{p} \big)^{\mathsf{H}}
\Big(
 \psi^{\text{opt}}_{sp}
-
 \text{diag}\big( \phi^{(k)}_{p}   \big)T_s^{(k)} 
\Big)
\bigg]
\\  
\phi^{(k+1)}_{p} &=  \label{supp:Eqn:seq_rPIE_update_phi}
\phi^{(k)}_{p}  +
\\
&\quad  \nonumber
\Big[
\text{diag}\Big( \big\vert T^{(k)}_{s} \big\vert^2 
+
w^{(k)}_{\sigma}  
\Big)
\Big]^{-1}
\Big[
 \text{diag}\big( T^{(k)}_{s} \big)^{\mathsf{H}}
\Big(
 \psi^{\text{opt}}_{sp}
-
 \text{diag}\big( T^{(k)}_{s}   \big) \phi^{(k)}_{p} 
\Big)
\Big],
\end{align}
where $k$ is update epoch index, $\textsf{H}$ is the conjugate transpose, and $\{ v_{\sigma}, w_{\sigma}\}$ are weights that control the steplength in the direction of steepest descent; they are defined as 
\begin{align}
v_{\sigma } &= \label{supp:Eqn:v_sigma}
\alpha_T \bigg[  \text{max}\Big( \sum_p \vert \phi_p \vert^2 \Big) 
- 
\sum_p 
 \vert \phi_p \vert^2
 \bigg]
\\  
w_{\sigma} &= \label{supp:Eqn:w_sigma}
\alpha_{\phi} \bigg[  \text{max}\big(  \vert T_s \vert^2 \big) 
-  
\vert T_s \vert^2
\bigg],
\end{align}
with $\alpha_{\phi} = 1$ and $\alpha_T \in [0, 1]$.

\subsection{Stochastic Minibatch PIE}
\noindent
The exit wave at scan position $s$ and for SCPM index $p$ can also be written as
\begin{align} \label{supp:Eqn:proj_approx_upper}
\Psi_{sp} = \text{diag}\left( \Phi_{sp} \right) T \in \mathds{C}^{M_T N_T \times 1},
\end{align}
where the quantity $\Psi_{sp}$ is simply the exit wave in array notation $\text{mat}\left( \psi_{sp}\right)$ given in Eq.~\ref{supp:Eqn:proj_approx_lower}, only zero padded both horizontally and vertically to be of the same size as the total sample array $\text{mat}\left(T\right) \in \mathds{C}^{M_T \times N_T}$. As before, we define the notation $\Psi_{sp}$ to mean that exit  wave is shifted to the scan position $s$ within the total sample array. 
From this, we make a straightforward modification to the traditional PIE updates rules in Sec.~\ref{supp:traditionalPIE} as follows:
\begin{align}  
 T^{(k+1)} &= \label{supp:Eqn:par_rPIE_update_T} 
T^{(k)} 
+
\\ 
&\quad \nonumber
\bigg[ \sum_{p, s \in \mathcal{B}_b} 
\text{diag}\Big( \big\vert \Phi^{(k)}_{sp}  \big\vert^2 
+
 v^{(k)}_{\pi} \Big)
 \bigg]^{-1}
\bigg[
\sum_{p, s \in \mathcal{B}_b} 
 \text{diag}\big( \Phi^{(k)}_{sp} \big)^{\mathsf{H}}
\Big(
 \Psi^{\text{opt}}_{sp}
-
\text{diag}\big( \Phi^{(k)}_{sp}   \big)T^{(k)} 
\Big)
\bigg]
\\  
\phi^{(k+1)}_{p} &=  \label{supp:Eqn:par_rPIE_update_phi}
\phi^{(k)}_{p} 
+
\\
&\quad \nonumber
\bigg[
\sum_{ s \in \mathcal{B}_b}
\text{diag}\Big( \big\vert T^{(k)}_{s} \big\vert^2 
+ 
w^{(k)}_{\pi}  
\Big)
\bigg]^{-1}
\bigg[
\sum_{ s \in \mathcal{B}_b}
 \text{diag}\big( T^{(k)}_{s} \big)^{\mathsf{H}}
\Big(
 \psi^{\text{opt}}_{sp}
-
 \text{diag}\big( T^{(k)}_{s}   \big) \phi^{(k)}_{p} 
\Big)
\bigg],
\end{align}
where $s \in \mathcal{B}_b$ means the sum over scan position index $s$ in only over our current batch $\mathcal{B}_b$, with $b \in \{ 1, 2, \cdots, N_b \}$ and $N_b$ as the total number of minibatches.
We call this \textit{stochastic} minibatch PIE because of the way these batches are chosen, for example if $s \in \{ 1, 2, 3, 4, 5\}$ then one possibility of minibatch selection would be $\mathcal{B}_1 \in \{ 5, 2, 4 \}$ and $\mathcal{B}_2 \in \{ 1, 3 \}$, with $N_b = 2$. We use a uniform random number generator to select out the scan position indices for the different minibatches using a ``sampling without replacement" scheme. We make a similar modification to the weights $\{ v_{\pi}, w_{\pi} \}$:
\begin{align}
v^{(k)}_{\pi}  &= \label{supp:Eqn:v_smb}
\alpha_T \bigg[ \text{max}\Big( \sum_{p, s \in \mathcal{B}_b } \vert \Phi^{(k)}_{sp} \vert^2 \Big) - 
\sum_{p, s \in \mathcal{B}_b }
 \vert \Phi^{(k)}_{sp} \vert^2
\bigg]
\\  
 w^{(k)}_{\pi}  &= \label{supp:Eqn:w_smb}
\alpha_{\phi} \bigg[ \text{max}\Big( \sum_{s \in \mathcal{B}_b }  \vert T^{(k)}_s \vert^2 \Big) 
- \sum_{s \in \mathcal{B}_b }   \vert T^{(k)}_s \vert^2
 \bigg].
\end{align}
Once the sample $T$ and SCPMs $\phi_p$ are updated according to 
Eq.~\ref{supp:Eqn:par_rPIE_update_T} and Eq.~\ref{supp:Eqn:par_rPIE_update_phi}, we then use further optimization methods \cite{Tripathi:14} for algorithmically updating the scan positions. 

We also use additional constraints on the ptychography problem to aid in convergence to higher quality local solutions.
The first constraint is that the SCPMs used to model partial spatial coherence properties of the probing wavefield must be mutually orthogonal, and this is accomplished as follows. We combine all of the SCPMs into a single matrix:
\begin{align}
\boldsymbol\phi
=
\begin{bmatrix} \phi_{1} & \phi_{2} & \cdots & \phi_{N_p} \end{bmatrix}
\in \mathds{C}^{M N \times N_p},
\end{align}
and then perform an eigendecomposition:
\begin{align}
{\boldsymbol\phi}^{\textsf{H}} \boldsymbol\phi
=
\mathbf{Q}^{\textsf{H}} \mathbf{D} \mathbf{Q}.
\end{align}
From this we obtain the orthogonalized SCPMs:
\begin{align}
\mathcal{P}_{\text{orthog}} [ \boldsymbol\phi ] =
\boldsymbol\phi \mathbf{Q}^{\textsf{H}},
\end{align}
where we also implicitly define the SCPM orthogonalization projection operator $\mathcal{P}_{\text{orthog}}$.
Another constraint for the probing wavefield is that we typically have a good estimation of the number of photons $N_{\text{photons}}$ incident on the sample at any given time, and we simply use this knowledge to define the constraint projection operator:
\begin{align}
\mathcal{P}_{\text{photons}  }[ \boldsymbol\phi ]
=\boldsymbol\phi
\sqrt{ \frac{ N_{\text{photons}}  }{ \sum_p \Vert \phi_p \Vert_2^2 }}.
\end{align}
For the sample constraints, we have the sample magnitude scaling constraint projection operator:
\begin{align}
\mathcal{P}_{\text{sample mag}}[ T ] = 
\begin{cases}
    T / \vert T \vert,                     
    & \text{for spatial regions of}~ \vert T \vert > 1
    \\
    T,       
    & \text{otherwise}.
\end{cases}
\end{align}
In words, this projection operator is implemented as: any sample pixels where $\vert T \vert > 1$ are set to be one while leaving the phase at those sample pixels unchanged, while for any sample pixels where $\vert T \vert \leq 1$ we leave the sample $T$ unchanged; this constraint is from the physical fact that from the definition of sample transfer function under the projection approximation \cite{paganin2013coherent} that the sample magnitude contrast is always positive and less than or equal to one.
Using these, the final update rules we use for the sample and SCPM reconstructions in this manuscript are:
 \begin{align}  
T_0^{(k+1)} &= \label{supp:Eqn:par_rPIE_update_T2} 
T^{(k)} 
+
\\ 
&\quad \nonumber
\bigg[ \sum_{p, s \in \mathcal{B}_b} 
\text{diag}\Big( \big\vert \Phi^{(k)}_{sp}  \big\vert^2 
+
 v^{(k)}_{\pi} \Big)
 \bigg]^{-1}
\bigg[
\sum_{p, s \in \mathcal{B}_b} 
 \text{diag}\big( \Phi^{(k)}_{sp} \big)^{\mathsf{H}}
\Big(
 \Psi^{\text{opt}}_{sp}
-
\text{diag}\big( \Phi^{(k)}_{sp}   \big)T^{(k)} 
\Big)
\bigg]
\\
 T^{(k+1)} &= \mathcal{P}_{\text{sample mag}}
 \left[ 
T_0^{(k+1)} 
 \right]
\\  
\phi^{(k+1)}_{p,0} &=  \label{supp:Eqn:par_rPIE_update_phi2}
\phi^{(k)}_{p} 
+
\\
&\quad \nonumber
\bigg[
\sum_{ s \in \mathcal{B}_b}
\text{diag}\Big( \big\vert T^{(k)}_{s} \big\vert^2 
+ 
w^{(k)}_{\pi}  
\Big)
\bigg]^{-1}
\bigg[
\sum_{ s \in \mathcal{B}_b}
 \text{diag}\big( T^{(k)}_{s} \big)^{\mathsf{H}}
\Big(
 \psi^{\text{opt}}_{sp}
-
 \text{diag}\big( T^{(k)}_{s}   \big) \phi^{(k)}_{p} 
\Big)
\bigg]
\\
\phi^{(k+1)}_{p} &= 
\mathcal{P}_{\text{photons}}
\left[
\mathcal{P}_{\text{orthog}}
 \left[ 
\phi^{(k+1)}_{p,0}
 \right]
 \right]
\end{align}

\newpage
\section{Reconstruction from bookshelf sample}
\label{supp:recon_bookshelf}

\begin{figure}[ht!]%
\centering
\includegraphics[width=1\textwidth]{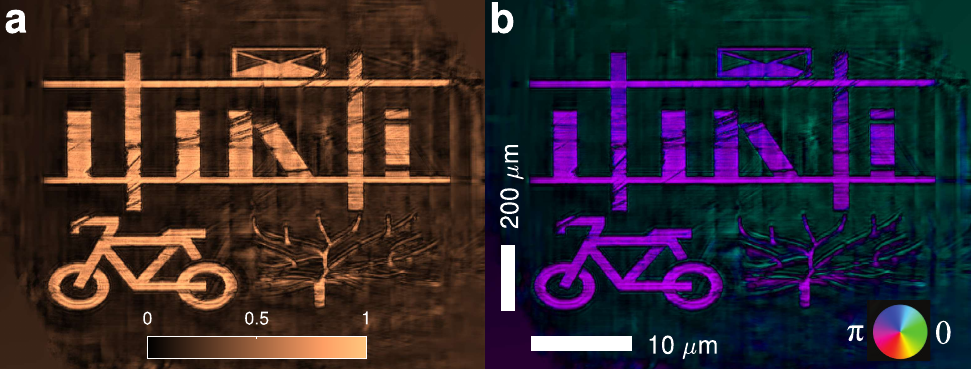}
\caption{(a) is reconstructed amplitude from the bookshelf sample of uniform height of 50~nm on silicon. (b) is the HSV representation (color for phase and brightness for amplitude). Data was taken at $1^\circ$ incident angle.}\label{supp:Fig:bookshelf}
\end{figure} 
\noindent
Ptychography from the bookshelf sample was collected using the same experimental conditions as from the double-Siemens sample. Identical reconstruction algorithm was employed for the reconstruction on all experiment samples, except that a low-resolution reconstruction (using a field of view of $256\times256$ on the detector) is used on bookshelf sample. It demonstrates that uniform height corresponds to uniform phases of the reconstructed transfer function.

\newpage
\section{Multi-slicing simulation for forward computations}
\label{supp:multislicing}

\begin{figure}[ht!]%
\centering
\includegraphics[width=0.8\textwidth]{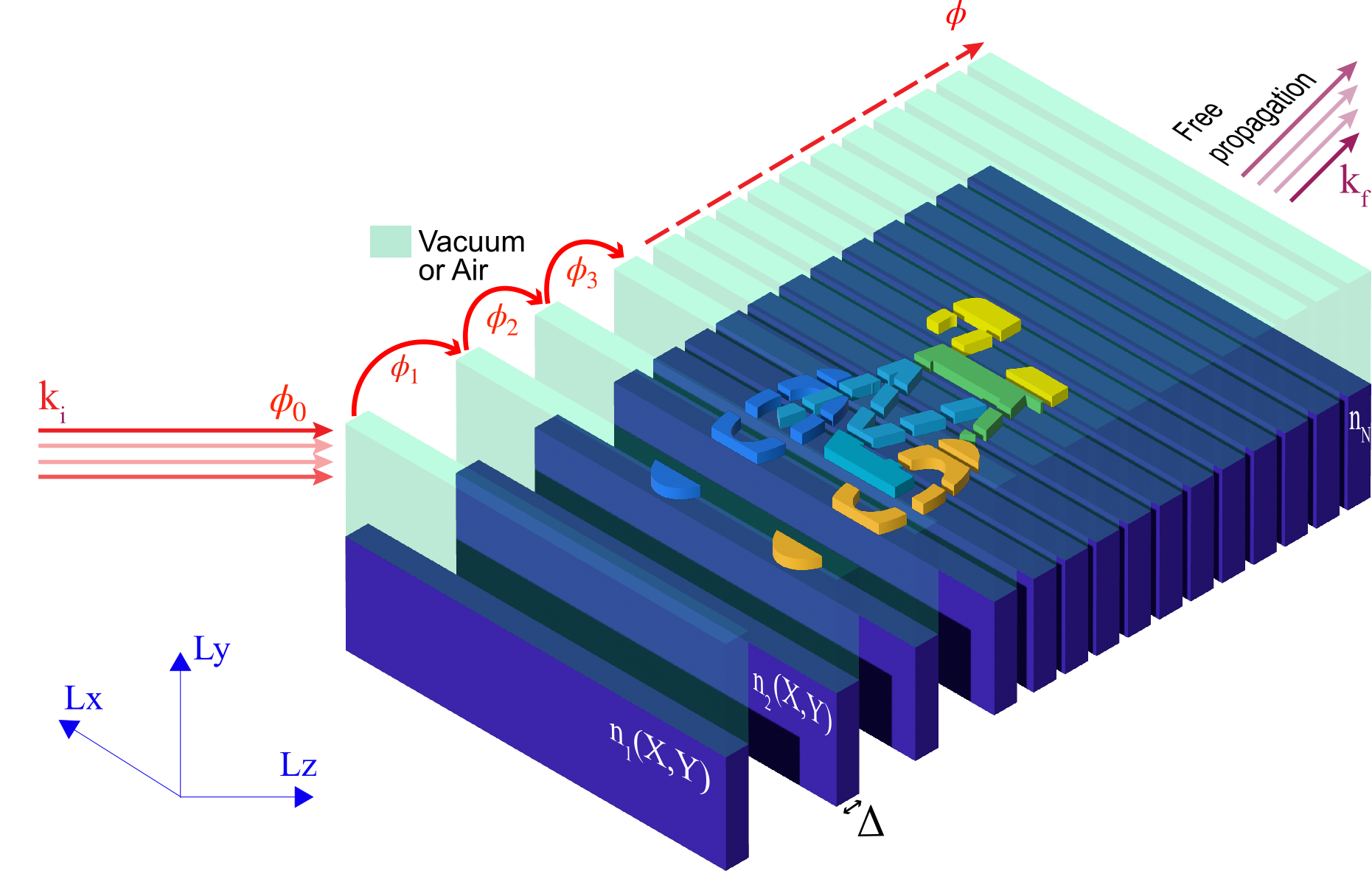}
\caption{Schematic diagram of multi-slicing wave propagation in the laboratory frame of reference for ptychography simulation. An incident wave of wave-front $\phi_0$ impinges on the first slice of thickness $\Delta$ and index of refraction $n_1(X,Y)$. Propagating through the first slice produces an exit wave $\phi_1$, which then becomes the incident wave for the second slice. This process continues until the last slice, after which a free propagation is performed to give the far-field scattering pattern on the detector.}\label{supp:Fig:multislicing}
\end{figure}
\noindent
DWBA faces a challenge when dealing with perturbation to the in-plane electric field due to structural inhomogeneities within the plane. This challenge becomes pronounced when a highly coherent beam interacts with anisotropic meso-scaled structures~\cite{chu2023}. Another constraint of DWBA is that it assumes a plane wave for the incident beam, which may not be readily met for many coherent imaging experiments. To accurately generate forward scattering for our simulation study, we used a multi-slicing Fresnel propagation method that has been previously developed for coherent surface scattering from nanostructures supported on substrates or buried in thin films~\cite{myint2023multislice}. As shown in Fig.~\ref{supp:Fig:multislicing}, three-dimensional patterns were first placed on top of a flat silicon substrate, and the entire object consisting of both pattern and substrate was rotated in the lab frame of reference according to the required incident angle. The rotated object was then sliced into two-dimensional field views, through which the incident wave propagates according to 
\begin{eqnarray}\label{supp:Eqn:multislice}
 \phi(X,Y,Z_0+\Delta)=&&\exp(ik\Delta) \mathcal{F}^{-1} \Bigg\{ \exp\Big(-i\Delta\frac{k_X^2+k_Y^2}{2k}\Big)\times \nonumber\\
                    &&\mathcal{F}\Bigg\{\exp\Big(\frac{k}{2i}  \int_{Z_0}^{Z_0+\Delta} [1-n^2(X,Y,Z)] \,dZ \Big)  \phi(X,Y,Z_0)\Bigg\}\Bigg\} \nonumber\\
\end{eqnarray}
where $k$, $n$, $\Delta$ and $\phi$ represent incident wave number, index of refection, slice thickness and the complex wave front. The exit wave of the previous slice becomes the incident wave of the next slice and this process iterates until the last slice, at which point free propagation is carried out to give the far-field scattering pattern observed on the detector. To simulate ptychography scans, the entire pattern was translated within the surface plane in the sample frame of reference. We used Fermat spiral sequence~\cite{huang2014optimization} to pre-compute scan positions in the lab frame and ensured on average there is an overlap of 70\% of the probed area (FWHM of the beam). These scan positions were then transformed to the sample frame for the pattern to translate correspondingly in the surface plane.  

The forward simulation was performed using a single NVIDIA A100 GPU with 80 GB memory. The X-ray energy and the photon flux were set to 7.35 keV and 5$\times10^{12}$ photons/s. A flat-top Gaussian beam with a full width at half maximum (FWHM) of 0.5~$\mu$m diameter is used as the incident probe. Voxel sizes for object discretization are 50~nm along the propagation direction $L_z$ (i.e. the incident beam direction), and 1~nm in the transverse $L_xL_y$ plane. The object volume for simulating each ptychographic scan location included the illuminated portion of the bicycle pattern, substrate, and space around the illuminated volume. For the incident angle range of $1^\circ$ to $1.9^\circ$ used in the simulation, dimensions of the volume used for the wave propagation were 40 $\mu$m along $L_z$, and $4\times4$~$\mu$m$^2$ in the transverse plane. The choice of 40~$\mu$m was well above the footprint on the sample surface for each scan location and each incident angle, ensuring distinct separation between the transmitted component (with respect to the real optical axis $OA_0$) and the reflected component (with respect to the virtual optical axis $OA_v$) in the exit wave from the last slice. The transmitted component was then discarded when calculating the far-field propagation (Fourier transform) for the scattering pattern on the detector. This approach is an analogy to the attenuation of the transmitted scattering channel in the substrate during grazing-incidence X-ray scattering. 

\newpage
\section{Dynamical scattering at grazing-incidence and distorted wave Born approximation}
\label{supp:dynamical_scat}

\begin{figure}[ht!]%
\centering
\includegraphics[width=0.9\textwidth]{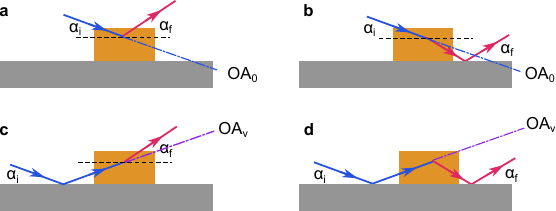}
\caption{Classical schematics to illustrate the four distinct scattering channels in grazing-incidence X-ray scattering. These channels correspond to the four terms in the effective form factor in the distorted wave Born approximation (DWBA) framework. (a) represents transmitted scattering channel with real incident beam. It is also referred as the kinematical approximation or Born approximation. (b) corresponds to reflected scattering channel with real incident beam. (c) is transmitted scattering channel with virtual incident beam, and (d) is reflected scattering channel with virtual incident beam. Scatterings from these four channels are categorized into to two groups: (a) and (b) is referred to the transmitted channel with respect to the optical axis $OA_0$ of the real incident beam, and (c) and (d) is the reflected channel with respect to the virtual optical axis $OA_v$.}\label{supp:Fig:DWBA}
\end{figure} 
\noindent
Distorted wave Born approximation (DWBA) is often used to calculate the forward scattering in the conventional grazing-incident X-ray scattering analysis~\cite{sinha1988x,renaud2009probing,jiang2011waveguide}. It reasonably well addresses multiple beam or dynamical scattering effects that are commonly observed in the grazing-incident or grazing-exit experiments. These dynamical effects arise from significant perturbations to the electric field along the surface normal direction caused by the supporting substrates. As a result, one cannot readily analyze surface scattering data using the single scattering assumption (kinematical approximation) that is widely used in most transmission scattering analyses and coherent diffractive imaging reconstructions. The principle of DWBA for a conventional grazing-incident X-ray scattering is illustrated in Fig.~\ref{supp:Fig:DWBA} for a simple example of a supported object on a substrate. The diffuse part of the differential scattering cross-section is written as the modulus square of the complex sum of four scattering channels:
\begin{equation}
\left(\frac{d \sigma}{d \Omega}\right)_{\text {diff}}=r_e^2|\Delta \rho|^2\left|\mathcal{F}\left(q_{\|}, k_z^i, k_z^f\right)\right|^2,
\end{equation}\label{supp:Eqn:DWBA_crosssection1} where $r_e$ is the classical electron radius, $\Delta\rho$ is the scattering contrast which is often the electron density contrast for hard X-rays, and $\mathcal{F}$ is the effective form factor of the object and is denoted by
\begin{align}\label{supp:Eqn:DWBA_crosssection2}
\mathcal{F}\left(q_{\|}, k_z^i, k_z^f\right)=&F\left(q_{\|}, q_z^1\right)+r_F\left(\alpha_f\right) F\left(q_{\|}, q_z^2\right)
+r_F\left(\alpha_i\right) F\left(q_{\|}, q_z^3\right)\\ \nonumber
&+r_F\left(\alpha_i\right) r_F\left(\alpha_f\right) F\left(q_{\|}, q_z^4\right),
\end{align}
where $q_{||}$ is the in-plane wave-vector transfer, $q_z^j$ ($j={1,2,3,4}$) is normal wave-vector transfer associated with each scattering channel, $F$ is the conventional form factor, a.k.a., Fourier transform of the object used in the kinematical approximation, and $r_F(\alpha_i)$ and $r_F(\alpha_f)$ are the Fresnel reflection and transmission coefficients, respectively, of the substrate.

It is worth emphasizing that in conventional grazing-incidence X-ray scattering, the specular scattering (i.e., reflectivity) and  diffuse scattering are typically treated as distinct components and are analyzed independently. The specular component is analyzed through reflectivity, while diffuse scattering is analyzed using the DWBA based formulas, such as ones described above. In the presence of a highly coherent incident beam on mesoscopic structures on surfaces, the distinction of specular and diffuse contributions, however, becomes less meaningful. In such cases, it is more advantageous to analyze the data using methods based on coherence analysis and coherent imaging reconstructions. Nevertheless, the discussion of the DWBA method in the conventional grazing-incidence X-ray scattering settings in here serves as an analogous representation to help understand the scattering contributions arising from combinations of reflected and transmitted scattering channels at various incident and exit angles, in order to illustrate the working conditions for ptychographic reflectometry imaging. 

The first term in Eq.~\ref{supp:Eqn:DWBA_crosssection2} corresponds to classical form factor given in the kinematical approximation (Fig.~\ref{supp:Fig:DWBA}a), which is the Fourier transform of the electron density of the object. The second term corresponds to the case where the scattered beam is reflected from the substrate with a complex weighting factor $r_F(\alpha_f)$ to account for the reflection on the exit side. Similarly, the third term represents scattering originating from a reflected incident beam (Fig.~\ref{supp:Fig:DWBA}c) with a weighting factor $r_F(\alpha_i)$. Lastly, the fourth term corresponds to the case where both the incident and scattered beam are reflected (Fig.~\ref{supp:Fig:DWBA}d). For multi-layered or buried structures, a self-consistent multi-layer DWBA approach is necessary to calculate the vertical perturbation to the electric field within each layer~\cite{jiang2011waveguide}. In the case of mesoscopic structures, a 3D finite-element DWBA can be used to model the perturbation to the in-plane electric field~\cite{chu2023}. While the above simple representation does not directly explain the origins of amplitude and phase contrast in ptychographic reflectometry imaging, it provides insight into the role of reflection and transmission components in coherent scattering in the surface geometry. For instance, the third and fourth terms in Eq.~\ref{supp:Eqn:DWBA_crosssection2} (Fig.~\ref{supp:Fig:DWBA}c and d) correspond to scattering with respect to the virtual optical axis $OA_v$, thus representing the reflected scattering channel. Conversely, the first and second terms (Fig.~\ref{supp:Fig:DWBA}a and b) relate to the real optical axis $OA_0$, representing transmitted scattering channel. In our simulation study and experimental setup, a high incident angle is used so that the reflected and transmitted scattering channels can be well separated so as to neglect the contributions from the first two terms. Moreover, the intensity of the Fresnel reflection decreases approximately as a power-law decay, specifically to the fourth power of the angle, when the exit angle is greater the critical angle. As a result, the fourth term becomes negligible in comparison to the third term when analyzing the surface scattering data using a high exit-angle region on the detector. In addition, the second term is further reduced. Therefore, surface scattering experiments with both incident and exit angle exceeding the critical angle effectively suppresses the dynamical scattering effects and distinctly separates reflected scattering channel from transmitted scattering channel. This strategy allows the ptychographic reconstruction from extended objects to use phase retrieval algorithms previously developed in the framework of the kinematical approximation, with only minor adaptations to account for the virtual probe and virtual optical axis.  

\newpage
\section{Penetration depth on silicon}
\label{supp:penetration_depth}

\begin{figure}[ht!]%
\centering
\includegraphics[width=1\textwidth]{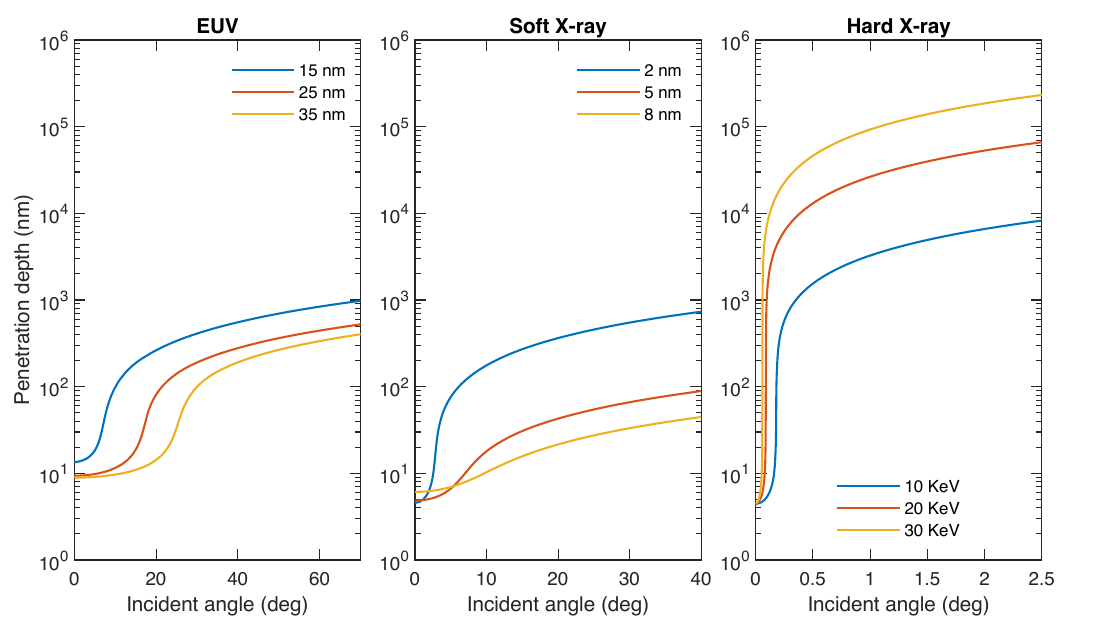}
\caption{Comparison of calculated penetration depths of EUV (wavelengths of 15, 25, and 35 nm), soft X-rays (wavelengths of 2, 5, 8 nm), and hard X-rays (energies of 10, 20, and 30 keV) as a function of incident angle on flat silicon substrates. Penetration depth is defined as the depth at which the electric field's amplitude diminishes to $1/e$ of the incident amplitude. Notably, hard X-rays has the capability to penetrate far beyond 1~$\mu$m into the material and offer a more extensive dynamic range of depth control through incident angle adjustment.}\label{supp:Fig:penetration}
\end{figure} 

\newpage
\section{Unwrapped relative phase angles of sample \#1 }
\label{supp:unwrapped_relative_phase}
\begin{figure}[ht!]%
\centering
\includegraphics[width=0.95\textwidth]{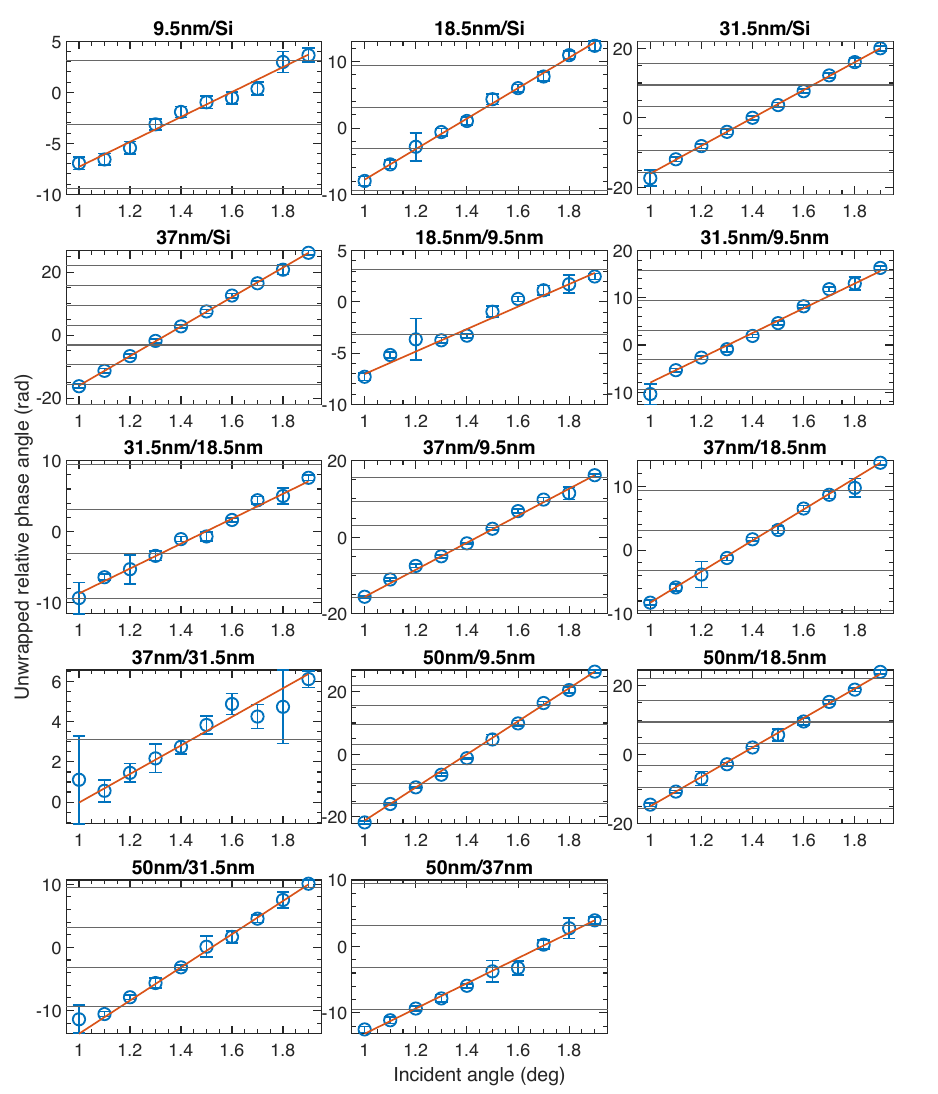}
\caption{Unwrapped relative phase angles of segment/substrate and segment/segment pairs (denoted in the panel titles) for sample \#1. Red lines are best weighted robust fits to the phase equation in the main text. Horizontal gray lines are phase angles $(2m+1)\pi$ ($m$ is an integer) where phases are wrapped in corresponding plots in the main text.}\label{supp:Fig:sample1_phase}
\end{figure} 

\newpage
\section{Result of sample \#2 of various material segments}
\label{supp:result_sam_2}
\begin{figure*}[ht!]%
\centering
\includegraphics[width=0.7\textwidth]{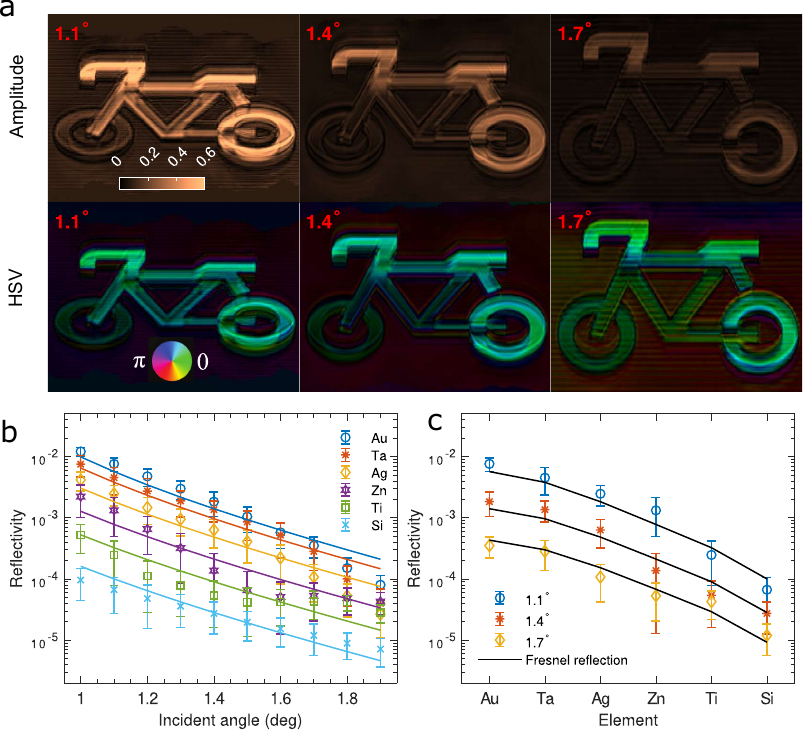}
\caption{Ptychography reconstruction from sample \#2 of various material segments. (a) shows the reconstructed amplitude (top row) and phase (HSV, bottom row) of the transfer function when viewed along the virtual optical axis for three incident angles: $1.1^\circ$,  $1.4^\circ$, and $1.7^\circ$. Averaged amplitude for each segment is plotted against the incident angle in (b) and against the material for three incident angles in (c). Solid lines in both (b) and (c) correspond to the calculated Fresnel reflectivity $|r_F(\alpha_i)|^2$.}\label{supp:Fig:sample2}
\end{figure*}
\noindent
Sample \#2 demonstrates how hard X-ray ptychographic reflectometry imaging can be sensitive to electron density contrast. The sample consists of segments of different elements, each with an identical height of 50 nm. At each incident angle, distinct segments exhibit varying amplitudes (top row in Fig.~\ref{supp:Fig:sample2}a). Notably, the Au segment exhibits the highest reflectivity, while the silicon substrate reflects the least. As the incident angle increases, the reconstructed amplitudes decrease for all segments. The ability to reflect the incident beam depends on the electron density of each segment, as shown by the shift in critical angle in Fig.~2d in the main text. This effect is also manifested as the vertical offsets of the reflectivity curves, which are dependent on the electron density (Fig.~\ref{supp:Fig:sample2}b and c). To accurately determine the heights of these segments, however, an angular sampling rate higher than $0.1^\circ$ increment is required, as discussed for sample \#1 in the main text. The reconstructed phase does not exhibit significant color variation across different segments at any angle (bottom row of Fig.~\ref{supp:Fig:sample2}a), except with the phase of the substrate. The uniform phases in segments is a result of their identical height and the absence of any variation in the path-length difference. This observation aligns with the expectation that phase contrast characterizes the surface topography. 

\newpage
\section{Wave vector transfers for grazing-incident small-angle geometry}
\label{supp:wave_vector_transfers}
\noindent
In the sample frame of reference (see Fig.~1 in the main text), the wave vector transfer for grazing-incidence geometry in each of the $(S_x,S_y,S_z)$ directions is given by 
\begin{align}
q_x & =k(\cos\alpha_f\cos2\theta-\cos\alpha_i) \label{supp:Eqn:qx} \\
q_y & =k\cos\alpha_f\sin2\theta  \label{supp:Eqn:qy} \\
q_z & =k(\sin\alpha_f+\sin\alpha_i).\label{supp:Eqn:qz}
\end{align}

The geometrical specular beam (GSB) refers to the intersection of the virtual optical axis ($OA_v$) with the detector and it is defined as the location on the detector where $\alpha_f=\alpha_i$ and $2\theta=0$ so that $q_x=q_y=0$. Unlike in traditional X-ray reflectivity and grazing-incident scattering, in a coherent surface scattering experiment, the most intense pixel on the detector may not correspond to the specular reflection. Accurately identifying GSB and the angle of $OA_v$ is critical for the imaging reconstruction, and this was done with high-precision alignment and calibration of the positions and orientations of both the sample and the detector. 

In the small-angle regime around GSB, $q_x$ can be written with the paraxial approximation for $OA_v$ as
\begin{equation}\label{supp:Eqn:qx_smallangle}
    q_x \approx -k\sin\alpha_i\sin(\alpha_f-\alpha_i).
\end{equation}
$q_x$ and $q_y$ are thus decoupled and become orthogonal near GSB as in the current study, eliminating the need of reciprocal space remapping due to the curvature of the Ewald sphere surface. Consequently, iterative phase retrieval algorithms originally developed for transmission geometry can be readily applied to our reflection geometry in the view of the virtual optical axis. A necessary adjustment is that the vertical dimensions of the reconstructed object need to be normalized by $\sin\alpha_i$ (as shown in Eq.~\ref{supp:Eqn:qx_smallangle}), in order to obtain the real dimension along the forward in-plane direction $S_x$ in the sample frame of reference. Similarly, one can scale the transverse resolution for $S_y$ from the Fourier ring correlation analysis to obtain the forward resolution along $S_x$. It is also noteworthy that the negative sign preceding the right side of Eq.~\ref{supp:Eqn:qx_smallangle} indicates that the reconstructed image is inverted up-side-down when viewed along the virtual optical axis. 

\bibliography{CSSIptycho}